\documentclass[
              reprint, twocolumn,
              preprintnumbers,
              amsmath,
              amssymb,
              showpacs,
              prl,
              superscriptaddress]{revtex4-1}

 \usepackage{ucs}
 \usepackage[utf8x]{inputenc}
 \usepackage{latexsym}
 \usepackage{amsfonts}
 \usepackage{amsmath}
 \usepackage[dvips]{graphicx}
 \usepackage{color}
 \usepackage{bm}
 \usepackage{hyperref}
 \usepackage{lineno}
 \usepackage[autostyle=false]{csquotes}
\graphicspath{ {./fig/} }
\hypersetup{
   colorlinks=true,       % false: boxed links; true: colored links
   linkcolor=magenta,          % color of internal links (change box color with linkbordercolor)
   citecolor=magenta,        % color of links to bibliography
   filecolor=magenta,      % color of file links
   urlcolor=blue           % color of external links
}
\begin{document}
%\linenumbers
\author{Krishna R. Nandipati}
\email[e-mail: ]{krishna.nandipati@pci.uni-heidelberg.de}
\affiliation{Theoretische Chemie,
             Physikalisch-Chemisches Institut,
             Universität Heidelberg,
             Im Neuenheimer Feld 229, 69120 Heidelberg, Germany}

\author{Oriol Vendrell}
\email[e-mail: ]{oriol.vendrell@uni-heidelberg.de}
\affiliation{Theoretische Chemie,
             Physikalisch-Chemisches Institut,
             Universität Heidelberg,
             Im Neuenheimer Feld 229, 69120 Heidelberg, Germany}
%\affiliation{Centre for Advanced Materials,
%        Universit\"at Heidelberg,
%        Im Neuenheimer Feld 205,
%       69120 Heidelberg, Germany}

%\author{coauthor 1}
%\email[e-mail: ]{coauthor@couathor.com}
%\affiliation{somewhere}

%\linenumbers

%\title{Jahn-Teller twist on chiral light inside an optical cavity}
%\title{Vibronic mixing of light polarizations in molecular polaritons}
%\title{Jahn-Teller twist on cavity photon polarization in molecular polaritons}
%\title{Cavity Jahn-Teller Polaritons of Molecules Interacting with Polarized Light}
%\title{Jahn-Teller splitting of molecular polaritons}
\title{Cavity Jahn-Teller Polaritons in Molecules}
\date{\today}

\begin{abstract}

%Molecular polaritonic states with mixed electronic, vibrational and photonic character promise a new handle to achieve cavity-controlled
%photophysics and photochemistry of
%single molecules and of molecular ensembles.
%
We investigate Jahn-Teller (JT) polaritons, which emerge
from the interaction of the two normal-incidence electromagnetic modes with perpendicular polarizations
in a Fabry-Perot cavity resonator with JT active systems.
These JT polaritons are characterized by a mixed $(+/-)$--circular electromagnetic polarization
that originates from the molecular JT vibronic coupling of the material subsystem.
%which mixes the degenerate
%electronic states coupled with the corresponding light polarizations.
%
Consequently, the exchange of photonic and vibronic angular momenta can be very efficient;
exciting the cavity-JT system with 
%exciting superpositions of different eigenstates supported by JT polaritonic potential energy surfaces (PESs) with
short, polarized light pulses
results in a dynamical and oscillatory response of the polarization
in the cavity medium.
Due to the photonic-vibronic coupling,
we show how the cavity polarization direction becomes frequency dependent and
does not necessarily coincide with the polarization direction
of the external fields used to drive the system.

\end{abstract}

\maketitle
%%%%%INTRODUCTION
%
%Recent use of optical cavities with confined light modes has proven quite promising for efficiently controlling various phenomena at molecular~\cite{hutchison2012modifying,schwartz2013polariton,morigi2007cavity,kowalewski2016non,galego2016suppressing,flick2017atoms,feist2018polaritonic,vendrell2018coherent} and material~\cite{ebbesen2016hybrid,kavokin2005optical,wang2019cavity} level, for instance, energy and charge transfer processes in organic/inorganic molecules and materials~\cite{ebbesen2016hybrid,ribeiro2018polariton,sun2017optical}. 
%
The formation of polaritons and identification of their light-matter composition are key to understanding cavity-controlled processes in molecules~\cite{hutchison2012modifying,schwartz2013polariton,herrera2020molecular,ribeiro2018polariton,kowalewski2016non,galego2016suppressing,flick2017atoms,feist2018polaritonic,vendrell2018coherent,herrera2016cavity,herrera2017dark,vendrell2018collective,dunkelberger2022vibration,morigi2007cavity} and materials~\cite{ebbesen2016hybrid,orgiu2015conductivity,wang2019cavity}. 
Polaritons with mixed electronic, vibrational and photonic character promise a new handle to achieve cavity-controlled photophysics and photochemistry of single molecules and of molecular
ensembles~\cite{herrera2016cavity,herrera2017dark,vendrell2018collective}.
%
%These vibronic polaritons, as we discuss, present non-trivial properties of the polarization degree of freedom of cavities.
%
%Various types of polaritons with properties that promise applications in electronics and photonics have been studied in molecules~\cite{herrera2020molecular} and materials~\cite{basov2016polaritons,grosso2017valley,xu2014spin,mak2012control,sun2017optical}. 
%
Recently, the possibility to control and exploit the photonic angular momentum
and the helicity of the cavity photons has received much attention~\cite{gautier2022planar,hubener2021engineering,feis2020helicity}.
Tuning the circular polarization of the polaritons can have profound implications in cavity-molecular
processes~\cite{hubener2021engineering,shelykh2009polariton}, for instance,
enantio-selectivity~\cite{yoo2015chiral,feis2020helicity} and polariton ring
currents~\cite{sun2022polariton}. 
%and optical spin-Hall effect~\cite{kavokin2005optical,wang2019cavity}. 
%
This has lead to the design of various schemes
that rely on the use of special mirrors and cavity configurations for controlling
the polarization and helicity of cavity modes~\cite{hubener2021engineering,feis2020helicity,abasahl2013coupling}.
While the mechanisms of vibronic coupling involving molecules and a single cavity mode
are well
understood~\cite{herrera2016cavity,herrera2017dark,vendrell2018collective,ribeiro2018polariton,dunkelberger2022vibration},
vibronic interactions involving
%Jahn-Teller (JT)-active systems with
circular cavity-modes
have, to the best of our knowledge, not
been addressed so far.
In highly symmetric molecular systems, electronic-state degeneracies can be lifted by
vibrational distortions of the molecular scaffold~\cite{bersuker2006jahn}. Hence the natural question arises,
to what extent do vibronic effects couple otherwise non-interacting
circularly-polarized cavity modes.
%cavity polarization directions. 

%the mechanisms of cavity polarization-direction mixing due to vibronic
%interactions in the material subsystem have, to the best of our knowledge, not
%been addressed so far.
%For instance, two opposite circular polarizations of a cavity mode, when coupled to molecular states representing two opposite angular momenta for electrons, can selectively interact to form corresponding circularly polarized polaritons. 
%
%If the electronic states with non-zero angular momentum are vibronically coupled, so are the circular polarizations of the cavity interacting with the molecules. This presents a non-trivial role of vibronic coupling in the formation of such circularly polarized polaritonic states.  
%

Molecular systems with an $n$-fold rotational symmetry axis $C_{n\geq
3}$ present doubly degenerate electronic states
belonging to two-dimensional irreducible representations of $E$ symmetry.
One of these $E$ subspaces always transforms according to the $x$ and $y$
functions in real space, the two components being labeled $E_x$ and $E_y$.
There are two properties of these $E$ electronic states that need to be
considered: first,  from simple selection rules, the two orthogonal polarization
directions of light propagating parallel to the $C_n$ symmetry axis couple the
totally symmetric $A$ ground electronic state of the molecules with the
corresponding $E_{x/y}$ component of electronically excited states. Examples are
found in the planar triazine or benzene molecules, featuring $C_3$ and $C_6$
rotation axes perpendicular to the molecular plane, respectively.   
Second, inevitably, the presence of vibronic coupling between $E_x$ and $E_y$
states in the molecules caused by displacements along vibrational modes creates
vibronic states with mixed electronic
character, i.e., the well-known Jahn-Teller (JT) effect~\cite{bersuker2006jahn,englman,Longuet-Higgins,koppel,whetten1986dynamic}.

%These JT coupled $E$ states, when coupled to a cavity mode with perpendicular polarizations yield the so-called JT polaritons (JTP).

\begin{figure}[t]
  \centering
 \includegraphics[width=1.0\columnwidth]{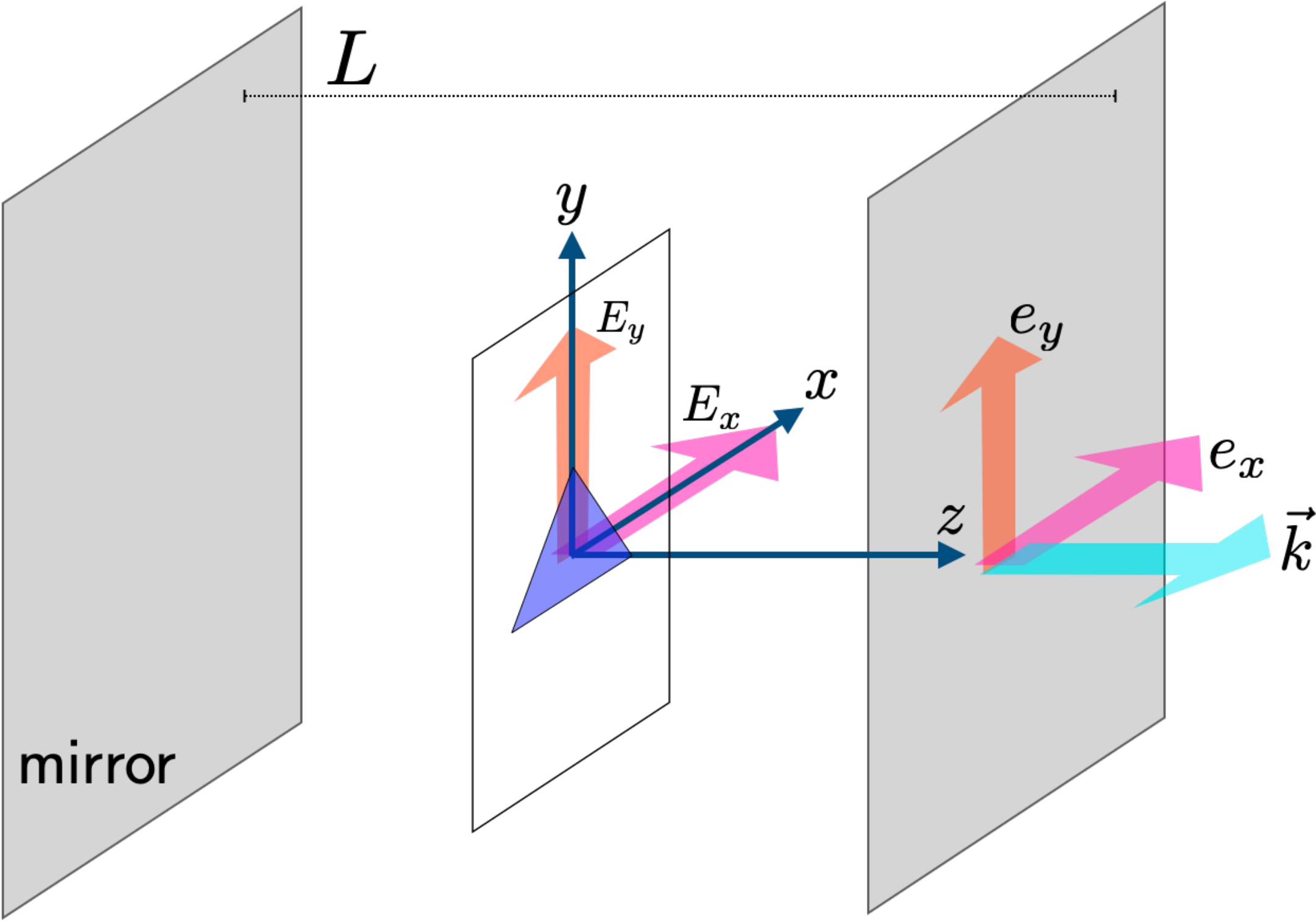}
  \caption{
  \label{fig:fig1}
     Schematic of a JT active system
     situated inside the FP cavity:
     The symmetry axis of the JT system, represented by the triangular shaped molecule,
     and the wave vector $\vec{k}$ of the $x(y)$-polarized cavity light point
     along the z-axis.
     The cavity is schematically represented
     by mirrors (in gray) separated by a distance $L$.
     The $x(y)$-polarized cavity mode interacts via dipole coupling with the $E_x$($E_y$)
     degenerate electronic excitation of the JT system.
     }
\end{figure}

We consider now that molecules with these characteristics lie with their $C_n$ axes perpendicular to
the mirrors of a
Fabry-Perot (FP) cavity resonator~(cf. Figure~\ref{fig:fig1}).
%Hence, if the molecules are planar,
%their plane lies parallel to the mirrors, although planarity is not required to form
%JTP, only the aforementioned rotational symmetry axis.
%
This configuration of the molecules and propagation direction of the FP electromagnetic modes
will result in the formation of JT polaritons, where the vibronically coupled
electronic states of the molecules couple with the two orthogonal polarization directions of the
normal-incidence cavity modes, thus mixing them.
The theoretical description of the JT polaritonic states, their mixed polarization character,
and the effect of the latter on the dynamics of the cavity polarization degree of freedom under external
perturbations, is the main subject of our investigation.

Without loss of generality, we base our description of the mechanism of
photonic-vibronic mixing of
the two cavity polarization directions
by considering the paradigmatic $(E\times e)$ JT Hamiltonian, where the $E$ electronic states
are coupled by the doubly degenerate $e$ vibrational modes~\cite{jahn1937stability,Longuet-Higgins,englman,bersuker2006jahn,koppel}.
The $(E\times e)$ coupling mechanism occurs, for example, in molecules with a $C_3$ symmetry axis.
The general properties of the $(E\times e)$ JT Hamiltonian have been well 
understood~\cite{moffitt1957configurational,Longuet-Higgins,sturge,englman,bersuker2006jahn}
in molecular spectroscopy as a premise to approach complex multi-mode vibronic interactions
in polyatomic systems~\cite{koppel}, thus making it an ideal model for investigating
molecular non-adiabatic effects with cavity modes.
%

% Discussion Figure 1

% %%%%% THEORY JT %%%%%%%%%%%%%%%%%%%%%%%%%%%%%%%%%%%%%%%%%%%%%%%%%%%%%%%%%%%%%%%%%%%%
\emph{Cavity Jahn-Teller Hamiltonian --}
The cavity-JT (CJT) Hamiltonian is comprised of
the JT molecular system plus the two normal-incidence modes of the FP
cavity with polarization directions $(\vec{e}_x, \vec{e}_y)$,
$\hat{H} = \hat{H}_{JT} + \hat{H}_C$, where $\hat{H}_{JT}$ is the linear
$(E\times e)$ JT Hamiltonian
and $\hat{H}_C$ is the cavity-Hamiltonian with light-matter coupling (see schematic in
Figure~\ref{fig:fig1}).
$\hat{H}_{JT}$ is represented in
the diabatic basis of the excited $E$ electronic states $|E_{x(y)}(\bm{Q}_0)\rangle$,
which are energetically well separated from the ground
electronic state $|A(\bm{Q}_0)\rangle$.
These electronic states are defined to be the eigenstates of
$\hat{H}_{JT} -\hat{T}_N$, the clamped-nuclei
Hamiltonian, at the reference nuclear geometry $\bm{Q}_0$.
For simplicity, in the following
we drop the indication $\bm{Q}_0$ inside the diabatic electronic state \textit{kets}.

In the electronic $\{|E_{+}\rangle, |A\rangle, |E_-\rangle\}$ basis,
the molecular JT Hamiltonian  reads
%\begin{align}\label{Hmol}
%    \hat{H}_{JT} 
%    & = \left(\hat{T}_N + \frac{\omega}{2} \rho^2\right) \mathbf{1}_{3\times 3} +
%        \left[
%    \begin{array}{ccc}
%       \epsilon             & 0        &  e^{-i\phi}   \\
%       0                    & 0        &  0                       \\
%       \kappa\rho e^{i\phi} & 0        &  \epsilon                
%     \end{array}
%     \right]
%\end{align}
%
\begin{align}\label{eq:Hmol}
    \hat{H}_{JT}
    & = \left(\hat{T}_N + \frac{\omega}{2} \rho^2\right) \mathbf{1}_{e} \\\nonumber
    & + \epsilon \left( |E_+\rangle\langle E_+| + |E_-\rangle\langle E_-| \right) \\\nonumber
    & + \kappa\, \rho\, \left( e^{i\phi} |E_-\rangle\langle E_+| + \text{h.c.}\right),
\end{align}
where $\hat{T}_N$ is the vibrational kinetic energy, 
$(\rho,\phi)$ correspond to the polar representation of the
$e$ vibrational modes, $\rho e^{i\phi} = Q_x + i Q_y$,
and $E_r$ refers to the complex basis
$|E_\pm\rangle = \left( |E_x\rangle \pm i |E_y\rangle \right)/\sqrt{2}$.
%and $\hat{b}^{(\dagger)}_r$ is the lowering (raising) operator for the
%$e$ vibrational mode $r$ \cite{Longuet-Higgins,koppel}. The latter operators
%are defined as
%$\hat{b}_\pm = (\hat{b}_x \mp i \hat{b}_y)/\sqrt{2}$  and  $\hat{b}_\pm^\dagger  = (\hat{b}_x ^\dagger \pm i \hat{b}_y^\dagger)/\sqrt{2}$.
%
$\mathbf{1}_{e}$ is the unit operator in the space of the electronic states,
$\omega$, $\epsilon$ and $\kappa$ are the
frequency of the $e$ modes, the energy of the $E$
electronic states at the reference geometry $\bm{Q}_0$,
and the linear
JT coupling parameter, respectively.
%
%% Vibronic angular momentum
%
This representation of $\hat{H}_{JT}$ makes particularly transparent that
the vibronic coupling results in an exchange
of angular momentum between the electronic subspace and the pseudo-rotation $\phi$
of
the molecular scaffold,
where the angular momentum perpendicular to the $(x,y)$-plane for the pseudo-rotation
of the vibrational modes is given as
\begin{align}
    %\hat{L}_z = i\hbar \left( \hat{b}_x \hat{b}_y^\dagger - \hat{b}_x^\dagger \hat{b}_y \right)
    %\hat{L}_z = \hbar \left( \hat{b}_+^\dagger \hat{b}_+ - \hat{b}_-^\dagger \hat{b}_- \right).
    \hat{L}_z = -i\hbar \frac{\partial}{\partial\phi},
\end{align}
and where one can introduce an electronic angular momentum-like operator within the $E$-subspace,
\begin{align}
   \hat{S}_z  = \hbar \left( |E_+\rangle\langle E_+| - |E_-\rangle\langle E_-| \right). 
\end{align}
The vibronic angular momentum of the molecular JT subsystem
is defined as $\hat{J}_\text{JT} = 2\hat{L}_z + \hat{S}_z$, where the factor $2$
in front of $\hat{L}_z$
follows from the $\pi$-radians periodicity of the vibrational pseudo-rotation instead of $2\pi$~\cite{Longuet-Higgins}.
It is a simple exercise to check that
%\begin{align}\label{CommJT}
    $
    [\hat{H}_{JT}, \hat{J}_\text{JT}] = 0,
    $
%\end{align}
the well-known symmetry of the linear $(E \times e)$ JT Hamiltonian
resulting in the double degeneracy of the vibronic spectrum. It is worth noting here
that the spectrum of the \emph{quadratic} JT Hamiltonian is also doubly degenerate due to the
remaining symmetry, although $\hat{J}_\text{JT}$ ceases to be a conserved quantity~\cite{Longuet-Higgins}.

The coupling of each cavity polarization to the corresponding electronic excitation is
now considered within the Condon approximation of constant transition dipole~\cite{tannor2007introduction},
and within the rotating wave approximation~\cite{cri_91_2430}
\begin{align}\label{eq:Hcav}
    \hat{H}_C & = \hbar\omega_c \left( \hat{a}_+^\dagger \hat{a}_+  + \hat{a}_-^\dagger \hat{a}_-\right)\\\nonumber
    & +
    \frac{\Omega}{2} 
    \left(\hat{a}_+^\dagger |A\rangle \langle E_+| + \hat{a}_-^\dagger |A\rangle \langle E_-| + \text{h.c.} \right).
\end{align}
%
%\begin{align}\label{Hcav}
%    \hat{H}_c = \sum_{r=x,y} \left[ \hbar\omega_c \hat{a}_r^\dagger %\hat{a}_r + \gamma  (\hat{a}_r^\dagger + \hat{a}_r) (|A_0\rangle %\langle E_r| + \text{h.c.}) \right]
%\end{align}
Here $\hbar\omega_c$ is the photon energy of the cavity modes, and $\Omega/2$ is
the coupling strength between the molecule and the cavity modes in energy units. Hence, at zero detuning the Rabi
splitting takes on the value $\Omega$ in units of energy.
For the representation of $\hat{H}_C$, we have introduced
circular cavity modes as
linear combinations of the $(x,y)$ linear polarizations,
$\hat{a}_\pm = (\hat{a}_x \mp i \hat{a}_y)/\sqrt{2}$  and  $\hat{a}_\pm^\dagger 
= (\hat{a}_x ^\dagger \pm i \hat{a}_y^\dagger)/\sqrt{2}$,
where $\hat{a}_\pm^{(\dagger)}$ annihilate (create) cavity photons with
$\pm\hbar$ angular momentum~\cite{cri_91_2430}.
Here 
we should be reminded that the $(+/-)$--circular modes of a cavity with conventional mirrors
have well-defined angular momentum (and thus a well-defined direction of rotation
of the electric and magnetic fields in the plane of the cavity as seen by
an external observer)
but have no helicity, the projection of the angular momentum of a particle
onto its linear momentum~\cite{gautier2022planar,hubener2021engineering}.
Thus, we will refer the angular momenta associated with these circular cavity modes in the following to the cavity \textit{polarizations}, but
we caution the reader that the cavity modes we are considering are not chiral.
%%%%%%%%%%%%%

%
%As we will see immediately, the CJTH spectrum is also doubly-degenerate
%as a consequence of the conservation of the total photonic plus
%vibronic 
%angular momentum with perpendicular projection.
%
After having introduced circular cavity modes, the photonic angular momentum of the FP cavity
perpendicular to the $(x,y)$
reads~\cite{cri_91_2430}
\begin{align}
    \hat{l}_z
             % & = i\hbar \left( \hat{a}_x \hat{a}_y^\dagger - \hat{a}_x^\dagger \hat{a}_y \right),\\ \nonumber
               & = \hbar \left( \hat{a}_+^\dagger \hat{a}_+ - \hat{a}_-^\dagger \hat{a}_- \right).
\end{align}
%where $\hat{a}_\pm = (\hat{a}_x \mp i \hat{a}_y)/\sqrt{2}$  and  $\hat{a}_\pm^\dagger  = (\hat{a}_x ^\dagger \pm i \hat{a}_y^\dagger)/\sqrt{2}$.
%
%Here, the operators $\hat{a}_{\pm}$/$\hat{a}_{\pm}^\dagger$ are defined analogously to the $\hat{b}_{\pm}$/$\hat{b}_{\pm}^\dagger$ above and they correspond to the circularly polarized photon modes.
%
%and the angular momentum of the cavity Hamiltonian~(\ref{Hcav}), i.e.
%including the electronic the cavity mode subspaces, which reads
%$\hat{J}_{c} = \hat{S}_z + \hat{l}_z$.
%
%It holds now that
%\begin{align}\label{CommJC}
%    $
%    [\hat{H}_c, \hat{J}_c] = 0
%    $, and
%\end{align}
Introducing the total vibronic-photonic angular momentum
$\hat{J} = 2\hat{L}_z + \hat{S}_z + \hat{l}_z$
and using the JT commutation relation introduced above
%Eqs.(~\ref{CommJT}) and (\ref{CommJC}),
one finds out that $\hat{J}$ is conserved for the CJT Hamiltonian,
%\begin{align} \label{CommTot}
    $
    [\hat{H}, \hat{J}] = 0.
    $
%\end{align}
%
It is straightforward to show that this commutation relation
holds as well for molecular ensembles, where
then $\hat{H} = \sum_{M} ( \hat{H}_{JT}^{(M)} + \hat{H}_C^{(M)} )$ and
$\hat{J}_\text{tot} = \sum_{M} ( 2\hat{L}_z^{(M)} + \hat{S}_z^{(M)} ) + \hat{l}_z$.
Thus, the vibronic-photonic eigenstates of the CJT Hamiltonian can be cast as eigenstates of the total
vibronic plus photonic angular momenta of the system and can be
characterized by the quantum number $j = 2L + s + l$. Here $L$ is
the angular quantum number of the vibrational pseudo-rotation,
$s=0, \pm 1$ represents the electronic angular momentum
of the $A$ and $E_\pm$
electronic states, respectively, and $l=n_+ - n_-$ is the angular momentum
quantum number of the cavity photons.

%They are not eigenstates
%of the photonic angular momentum $\hat{l}_z$
%due to the coupling with the vibronic subsystem,
%and therefore
%
%they have a mixed cavity polarization
%character. Even if higher orders of the JT Hamiltonian are
%considered, the doubly degeneracy remains, as well as the mixing between the
%components of the individual angular momenta.

% Discussion Figure 2
\begin{figure}[t]
  \centering
  \includegraphics[width=0.9\columnwidth]{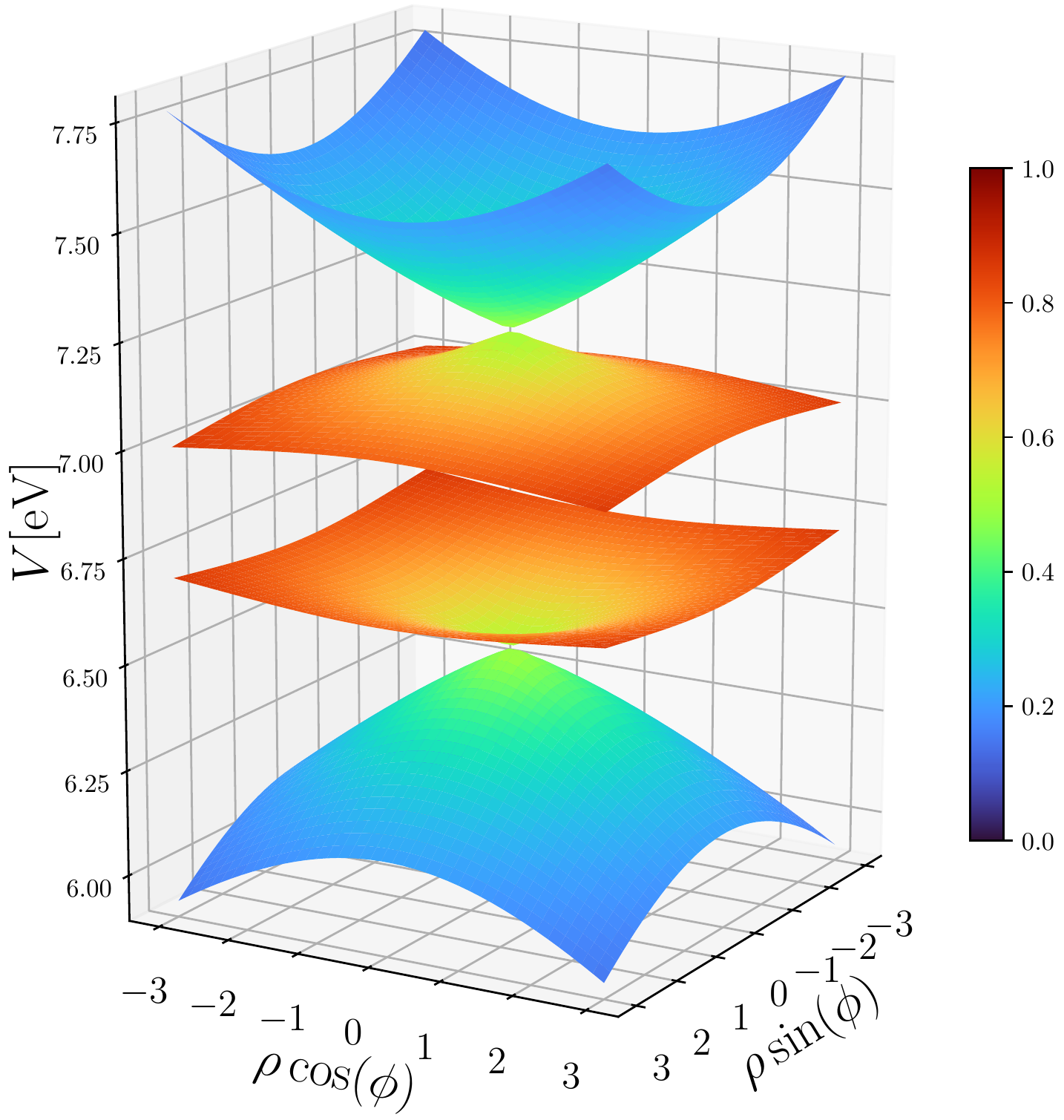}
  \caption{JT polaritonic PESs as a function of the $e$ normal mode coordinates (dimensionless), $x=\rho cos(\phi)$ and $y =\rho sin(\phi)$, plotted for $\Delta$ =0 and $\Omega = 0.75$~eV. The PESs are color-coded with the molecular and photonic contributions at a given $(x,y)-$The color bar on the right shows the
    photonic contribution on a 0 to 1 scale, where 1 indicates 100\%
    of photonic character.
  \label{fig:fig2}
     }
\end{figure}
%
% The electronic (and vibronic)-polaritons ---
\emph{Jahn-Teller Polaritons--} 
We are primarily interested in describing the strong light-matter
coupling regime, where light-matter coupling dominates over the vibronic non-adiabatic coupling, i.e. the Rabi splitting $\Omega$ is at least of the order of the width of the JT spectrum of the bare molecule.
Hence, we introduce the basis
of polaritonic states of the electronic-photonic subsystem (in the single-excitation subspace),
\begin{align} \label{eq:polaritons}
    |\Lambda_\pm^{(+1)}\rangle & =
    \left(|A, 1_{+}, 0_-\rangle \pm |E_{+}, 0_{+}, 0_-\rangle
    \right)/\sqrt{2}\\\nonumber
    |\Lambda_\pm^{(-1)}\rangle & =
    \left(|A, 0_{+}, 1_-\rangle \pm |E_{-}, 0_{+}, 0_-\rangle \right)/\sqrt{2}
\end{align}
The superscript $p=\pm 1$ in $|\Lambda_\pm^{(p)}\rangle$ indicates the combined electronic-photonic
angular momentum, i.e. $p = s + l$.
The subscript $\pm$ indicates the upper/lower polaritonic state.
In its matrix representation and in the polaritonic states basis (in the given order)
  $\{|\Lambda_+^{(+1)}\rangle, |\Lambda_+^{(-1)}\rangle,
     |A, 0_+, 0_-\rangle,
     |\Lambda_-^{(+1)}\rangle, |\Lambda_-^{(-1)}\rangle\}$,
the CJT Hamiltonian reads
\begin{widetext}
\begin{align} \label{eq:CJTH}
\hat{H} =
    \left( \hat{T}_N + \frac{\omega}{2} \rho^2\right) \mathbf{1}_{5\times 5}
 + \frac{\kappa \rho}{2}
 \left[
    \begin{array}{ccccc}
       0                         & e^{-i\phi}      & 0         &  0                  & -e^{-i\phi}  \\
      e^{i\phi}                & 0                & 0         & -e^{i\phi}         &  0           \\
      0 & 0 & 0 & 0 & 0 \\
       0                         & -e^{-i\phi}      & 0         &  0                  & e^{-i\phi}  \\
      -e^{i\phi}                & 0                & 0         & e^{i\phi}         &  0           \\
     \end{array}
     \right] 
    +  \left[
    \begin{array}{ccccc}
       \Sigma + \Omega/2         & 0                 & 0         &  \Delta             & 0            \\
       0                         & \Sigma + \Omega/2 & 0         &  0                  & \Delta       \\
       0 & 0 & 0 & 0 & 0 \\
       \Delta                    & 0                 & 0         &  \Sigma - \Omega/2  & 0            \\
       0                         & \Delta            & 0         &  0                  &  \Sigma - \Omega/2   \\
     \end{array}
     \right].
\end{align}
\end{widetext}
The second term in in Eq.~(\ref{eq:CJTH}) decribes the vibronic coupling between
the two upper and the two lower polaritonic basis states.
The upper-left $2\times 2$ submatrix
describes the vibronic mixing of the upper $|\Lambda_+^{(\pm 1)}\rangle$,
whereas the lower-right $2\times 2$ submatrix describes the vibronic mixing of the lower
$|\Lambda_-^{(\pm 1)}\rangle$ polaritonic states, respectively.
The third term corresponds to the matrix representation of $\hat{H}_C$ (cf. Eq.~\ref{eq:Hcav}),
where $\Sigma = (\hbar\omega_c + \epsilon)/2$, and
$\Delta = (\hbar\omega_c - \epsilon)/2$ is half the cavity detuning.
%and $\Omega$ corresponds to the Rabi splitting at zero detuning.
%

The diagonalization of the clamped-nuclei CJT Hamiltonian, $\hat{H}-\hat{T}_N$,
as a function of the vibrational displacements $(\rho,\phi)$ results in the coupled
JT polaritonic potential energy surfaces (PESs), which we refer to henceforth as
JT polaritons, and
which are represented in Fig.~\ref{fig:fig2} for $\Delta=0$ and $\Omega=0.75$~eV.
The two upper JT polaritonic surfaces are connected by a conical intersection (CI) that has been inherited from
their molecular contribution. The coloring of the JT polaritonic surfaces indicates their photonic (red) and
molecular (blue) contributions, with the strongest light-matter mixing
at the CI.
Likewise, a CI connects the two lower JT polaritonic surfaces.
This picture makes it clear that the molecular non-adiabatic coupling is
ultimately responsible for the mixing of cavity photons with positive and negative angular momentum (or circular polarization).

\emph{Spectrum and properties of the cavity Jahn-Teller Hamiltonian --}
Diagonalization of the CJT Hamiltonian $\hat{H}$ (cf. Eq.~(\ref{eq:CJTH}))
yields the doubly degenerate eigenstates supported by JTP
belonging to $j=\pm 1$ blocks:
%The $\hat{H}_{JT}$ with the inclusion of the bare vibrational and cavity parts of the CJTH $\hat{H}$, which  upon diagonalization in the harmonic basis of vibrational and cavity modes, yields the doubly degenerate JTP states with $\pm J_{tot}$, $|j_{J_{tot}}\rangle$, which can be written as (we fix $|J_{tot}|$ = 1 in what follows)

%The eigenstates of the CJTH, \{$|j\rangle$\}, can be expressed conveniently in the diabatic polariton basis which, within the single-excitation subspace of the cavity-molecule interaction and within the space of diabatic electronic states \{$A_0$,$E$\}, is given by $|\Lambda_\pm^{x(y)}\rangle$ = $\left(|A_0, 1_c\rangle \pm |E_{x(y)}, 0_c\rangle \right)/\sqrt{2}$, where $|A_0, 1_c\rangle$ represents the JT molecule in its ground state with one x(y)-polarized cavity photon and $|E_{x(y)}, 0_c\rangle$ represents the molecule in the excited state in the cavity vacuum. Owing to the JT coupling of the component $E_x$ and $E_y$ electronic states by $e$ vibrational modes, the electronic polaritons with independent polarizations become mixed, leading to the JT polaritonic states, $|j\rangle$, with mixed polarization (cf. Fig.~\ref{EJT_spec}(b)):
\begin{align}
    \label{eq:jth}
     |k_{j}\rangle
     & =
    \sum_{p,r=+,-}|\Xi_{p,r}^{(k_j)}\rangle
    \otimes
    |\Lambda_r^{(p)}\rangle,
    %+
    %|\Xi_{-,(r)}^{(j_{\pm 1)}}\rangle
    %\otimes
    %|\Lambda_-^{(r)}\rangle)\right]
\end{align}
%where $|\Xi_{\pm,x(y)}^{(j)}\rangle$
where $|\Xi_{p,r}^{(k_j)}\rangle$
is the vibrational contribution (since the total electronic and photonic angular momentum $p$ can
only take the values $\pm 1$, we refer to it as $\pm$ henceforth).
%$\langle\Xi_{s,(r)}^{(j_{\pm 1})}|\Xi_{s,(r)}^{(j_{\pm 1})}\rangle$ gives the population of the component $|\Lambda_s^{(r)}\rangle$, $p$. The coefficients $|\Xi_{s,(r)}^{(j_{\pm 1})}\rangle$ vary solely with the strength of the JT coupling $\kappa$ which determines the degree of cavity polarization presented by each $j_\pm 1$. 
%
%Clearly, the JT coupling mixes the $x$ and $y$-polarized (or equivalently, left and right circularly polarized) polaritons [cf. Fig.~\ref{JTC_coupling_spec}(a)].
%
The expectation value of the angular momentum of the cavity photons
in the eigenstate $|k_j\rangle$ is given by
\begin{align}
    \label{eq:pops}
    \mathcal{P}_{k}^{(j)} = \langle k_j| \hat{l}_z| k_j \rangle & =
    \langle \Xi_{+,+}^{(k_j)} |  \Xi_{+,+}^{(k_j)} \rangle +
    \langle \Xi_{+,-}^{(k_j)} |  \Xi_{+,-}^{(k_j)} \rangle \\\nonumber
    & -
    \langle \Xi_{-,+}^{(k_j)} |  \Xi_{-,+}^{(k_j)} \rangle -
    \langle \Xi_{-,-}^{(k_j)} |  \Xi_{-,-}^{(k_j)} \rangle.
\end{align}
%Analogously, the $\langle j_{\pm 1}|\hat{S}_z|j_{\pm 1}\rangle$ quantifies the net electronic angular momentum of the molecule in the state $j_{\pm 1}$, which is directly related to the electronic ring current ($C_{j_{\pm 1}}$) via $\hat{C}_{j_{\pm 1}} = \hat{S}_z/m_ea_0^2$ (where $m_e$ is the mass of the electron and $a_0$ is the Bohr radius)~\cite{nandipati2021dynamical}. 
By symmetry, $\mathcal{P}_k^{(1)} = -\mathcal{P}_k^{(-1)}$.
%$\mathcal{C}_{j_{+1}} = -\mathcal{C}_{j_{-1}}$ 
%
%By taking the $|j\rangle$ in Eq.~(\ref{eq:jth}) in the complex representation, one can straightforwardly show that the quantity $\langle j|\hat{l}_z|j\rangle$ is the population imbalance of the ground $A_0$ state associated with the left and right circular cavity photon whereas the quantity $\langle j|\hat{S}_z|j\rangle$ gives the population imbalance of the excited $E$ component states associated with the cavity vacuum. If $|j\rangle$ becomes time-dependent by means of external perturbation, so are the net polarization and the current. In what follows, for comparison purposes, we discuss also the ring currents along with the polarizations presented by the vibronic polariton state $j$.
%
%#######################################################################################################
%%%RESULTS&DISCUSSION%%%%%%%%%%%%%%%%%%%%%%%%%%%%%%%%%%%%%%%%%%%%%%%%%%%%%%%%%%%%%%%%%%%%%%
\begin{figure}[t]
  \centering
 \includegraphics[width=0.95\columnwidth]{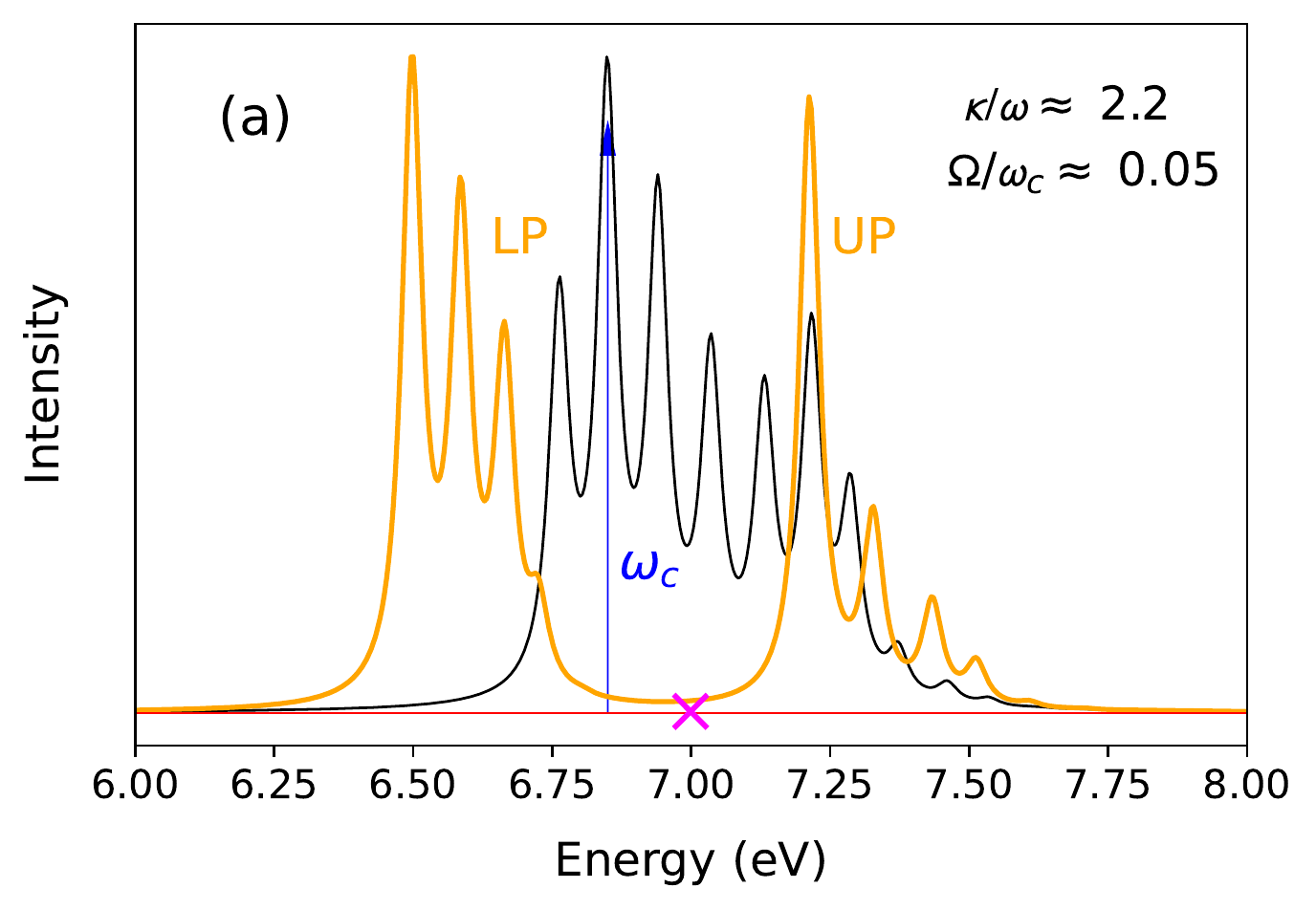}\\
 \vspace{0.5cm}
 \includegraphics[width=0.95\columnwidth]{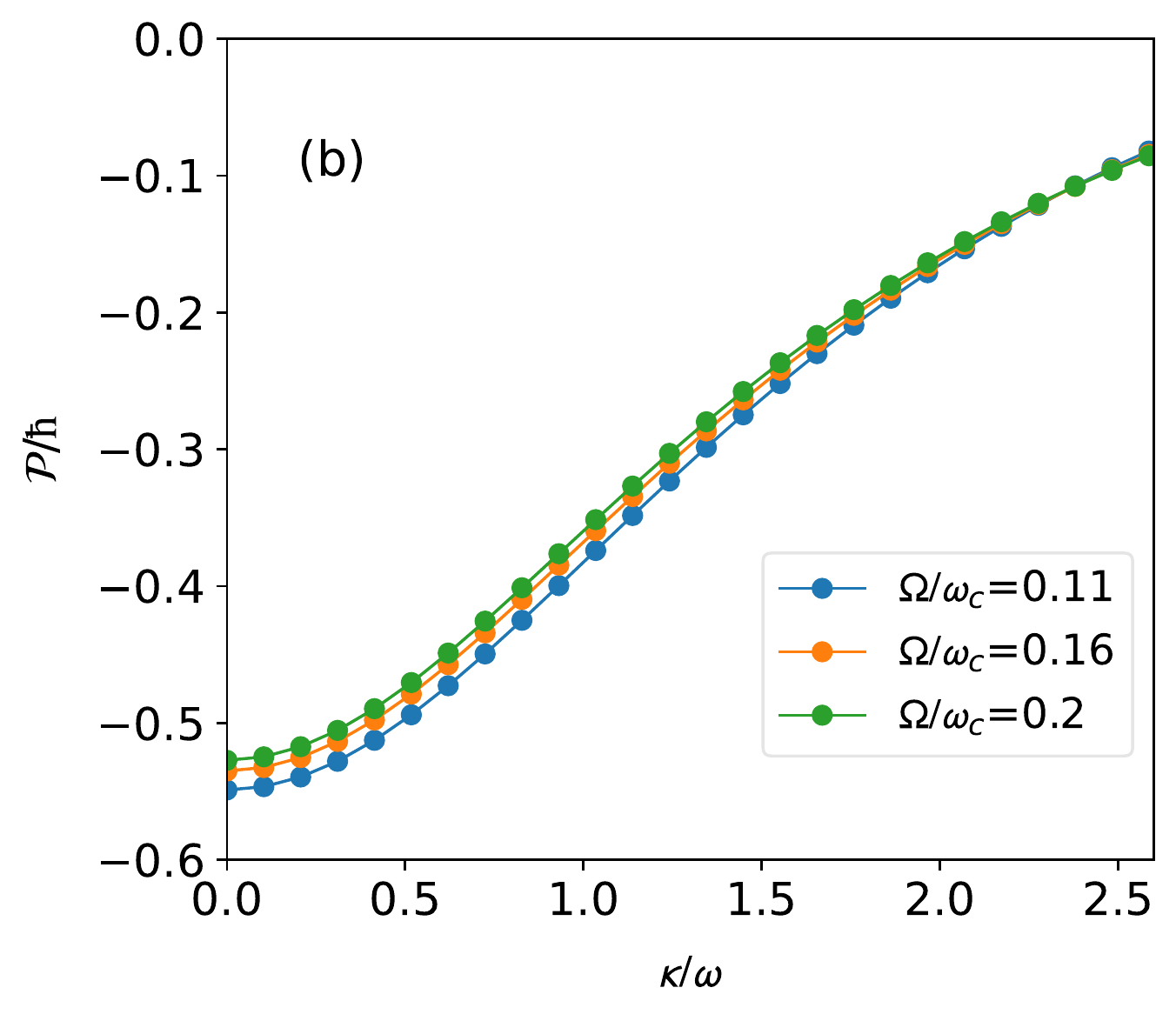}
 \caption{
  \label{CEJT_spec_pol}(a) The spectrum of the ($E\times e$) JT model of sym-triazine (in black) and
  the spectrum of the same system coupled to the FP cavity (in orange) in Fig.~\ref{fig:fig1}. The $\omega_c$ is set at the energy of the most optically bright state of the bare JT spectrum (indicated with vertical blue line) of the triazine, which is lying below the CI energy $\epsilon$ (marked \textcolor{magenta}{$\times$} on the energy axis).
%$\omega_c$ is set the energy of the most optically bright state of the bare JT spectrum (indicated in vertical blue line). 
(b) The net polarizations presented by the most optically bright $k$ state within $j=-1$
block as a function of the JT coupling ($\kappa/\omega$), at selected cavity-molecule couplings ($\Omega/\omega_c$).
}
\end{figure}
The cavity angular momentum (or cavity polarization) $\mathcal{P}_k^{(j)}$ is strongly dependent on the
JT coupling, which determines the mixing of both polarization directions in
each eigenstate.
%These polarizations/currents are then generated by external CP pulses. 
%We mention that the numerical spectrum of the EJT system  the dipole-selection rule involved is with respect to the cavity photon rather than the JT molecule because the former has more absorbance than the latter in reality. 
%
We illustrate the dependency of the cavity polarization on a vibronic coupling model with parameters
$\omega=0.003$~a.u. ($\approx 660$~cm$^{-1}$) and $\epsilon$ = 7~eV (cf. Eq.~\ref{eq:Hmol})
which are in the typical range for JT active vibrational modes and vertical electronic transitions
organic molecules such as benzene ($D_{6h}$) \cite{worth2007model} and sym-triazine ($D_{3h}$)~\cite{whetten1986dynamic}. 

Figure~\ref{CEJT_spec_pol}a presents the spectrum of the CJT model of sym-triazine with the JT coupling, $\kappa/\omega$ = 2.2~\cite{whetten1986dynamic} and the cavity coupling, $\Omega/\omega_c$ = 0.05. 
%The $\omega_c$ is set at the energy of the most optically bright state of the bare JT spectrum (indicated in blue) of the triazine, which is lying below the CI energy $\epsilon$ (marked \textcolor{magenta}{$\times$} on the energy axis). 
%
%The states in the CJT polaritonic spectrum (in orange) are doubly degenerate, characterised by  $J=\pm 1$, and 
The states in the spectrum are doubly degenerate, characterised by  $j=\pm 1$ and the spectrum splits into lower polariton (LP) and upper polariton (UP) branches that are separated by $\Omega$.
%The  $\omega_c$ (cf. Eq.~\ref{Hcav}) is set equal to $\Delta$.
%
The photonic angular momentum $\mathcal{P}_k^{(-1)}$ for the most optically bright $k_{-1}$ state as a function
of JT coupling ($\kappa/\omega$), at different cavity-molecule couplings ($\Omega/\omega_c$ $\geq$ 0.5 eV) are shown in Figure~\ref{CEJT_spec_pol}b.
% The $P$ are plotted, respectively, in $\hbar$  units. 
%
The magnitude of the cavity polarization $|\mathcal{P}|$ is strongly affected by the JT coupling $\kappa/\omega$. 
%
%Both $|P|$ and $|C|$ decrease from the maximum initial value of 1.0 (normalized, when $\kappa/\omega \to 0$) with increasing $\kappa/\omega$, for various values of $\gamma/\omega_c$, in a similar fashion. 
%
%
%The difference in $|P|$ and $|C|$ is owing to the different vibrational contributions associated with the $|A_0\rangle$ (with one left/right circular photon) and $|E\rangle$ states (with no photon) in the vibronic polariton state $j$ (cf. Eq. (~\ref{eq:jth})). 
%
At strong vibronic couplings, \mbox{$\kappa/\omega$ $>1$}, the net polarization in the cavity is almost suppressed to 10\%, even at the strongest cavity coupling $\Omega/\omega_c$. 
Note that $|\mathcal{P}|$ for $\kappa=0$ is determined by the cavity-molecule detuning parameter $\Delta$ (here $2\Delta$ = 0.15 a.u.).
%The effect of $\Omega/\omega_c$ on $P$ is evidently marginal compared to that of $\kappa/\omega$ which points to the fact that vibronic coupling is the key in determining the cavity photon polarization.

%, as is for the electronic ring currents. 
%\begin{figure}[t]
%  \centering
% \includegraphics[width=0.95\columnwidth]{fig3.pdf}
%  \caption{
%  \label{EJT_polcurr}
% The net polarizations (a) and the net currents (b) presented by the most optically bright $j$ state within $J_{tot}$ = -1 block as a function of the JT coupling ($\kappa/\omega$), at selected cavity-molecule couplings ($\gamma/\omega_c$).
%     }
%\end{figure}
%

% Discussion Figure 2
%
 \begin{figure}[!t]
  \centering
  \includegraphics[width=0.85\columnwidth]{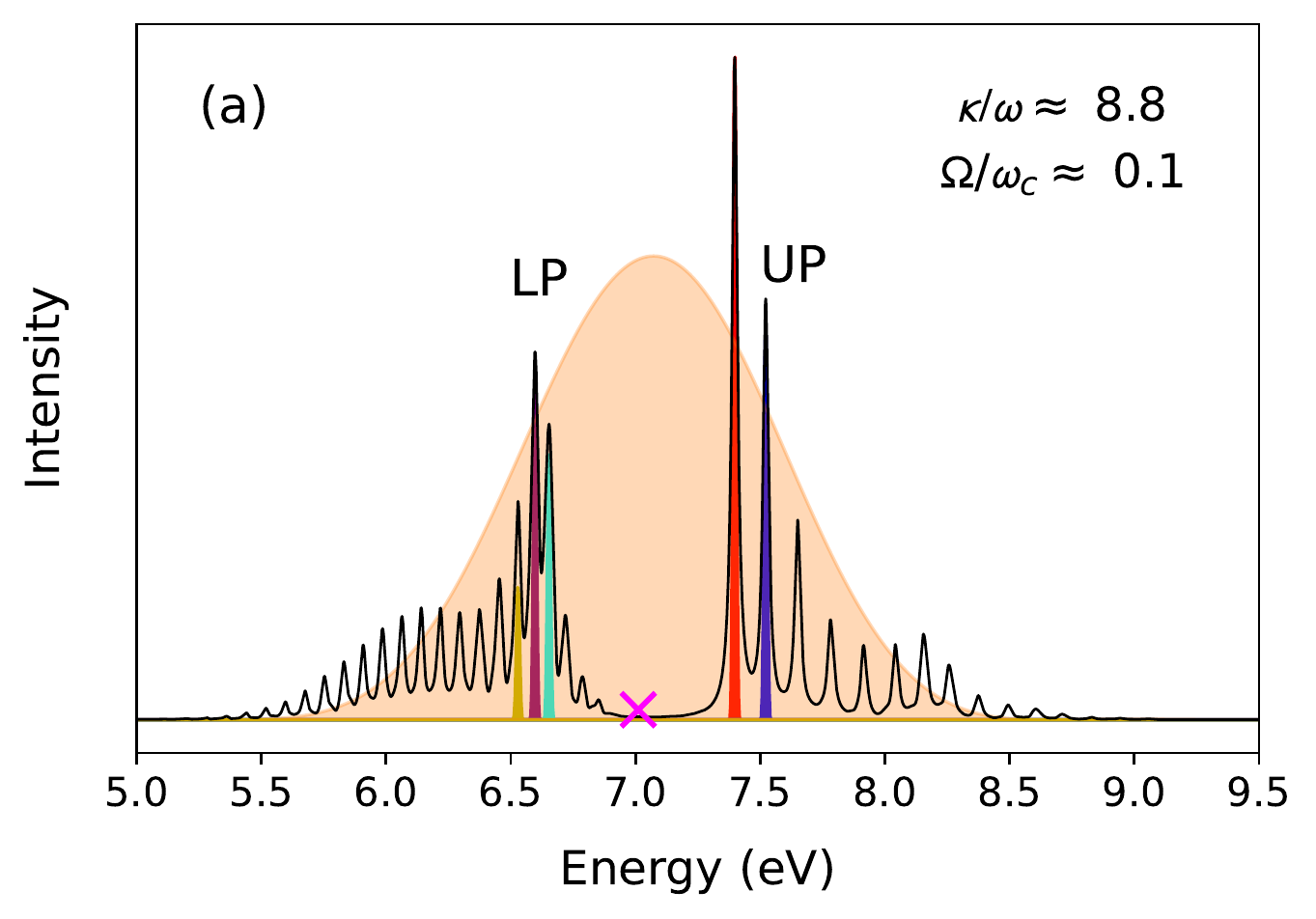} \\
  %\vspace{-0.3cm}
  \includegraphics[width=0.95\columnwidth]{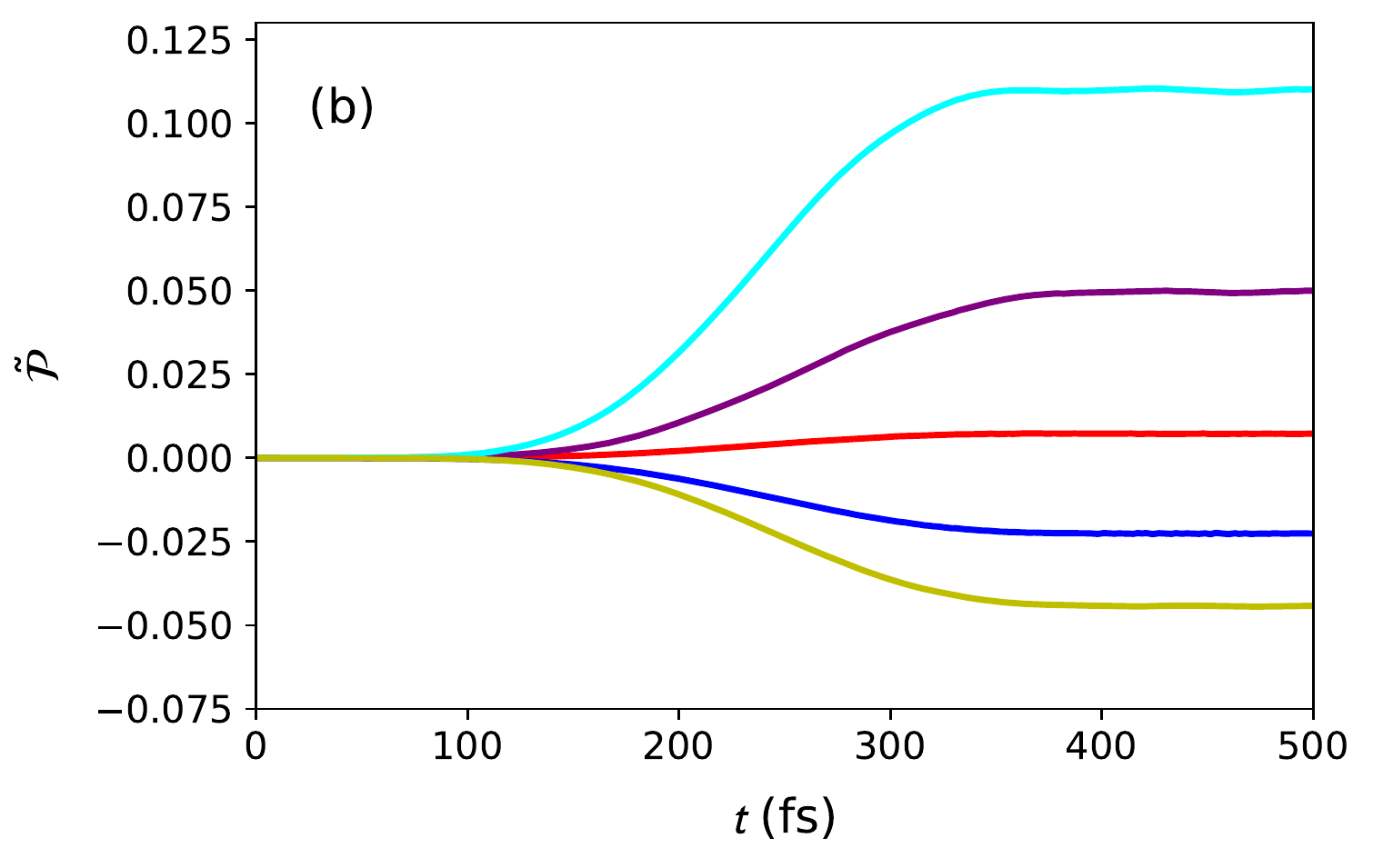} \\
  %\vspace{-0.3cm}
  \includegraphics[width=0.95\columnwidth]{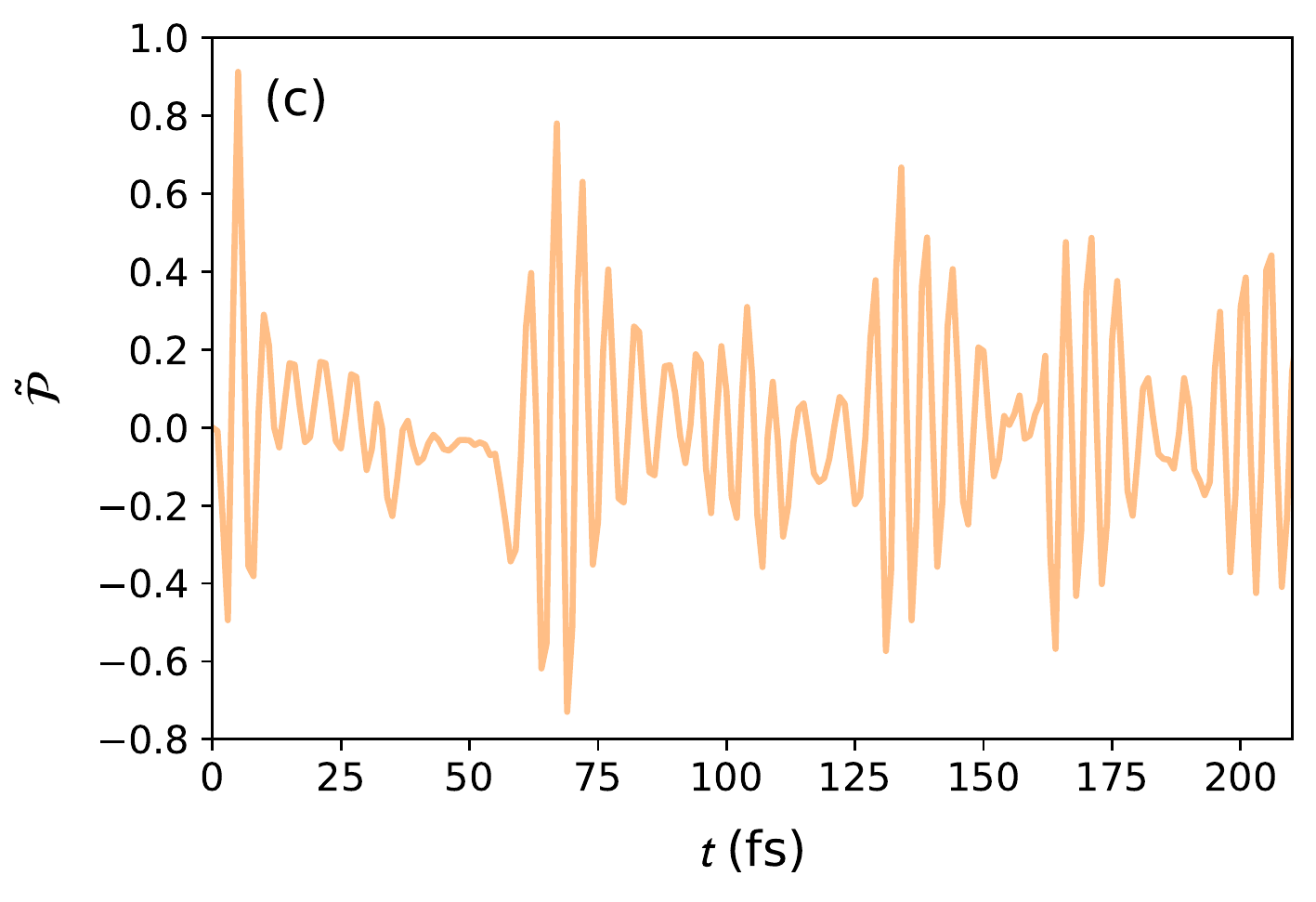}
  \caption{
  \label{el_ph_Imbalance_gen}
     (a) The spectrum of the CJT system
     for  strong JT coupling $(\kappa/\omega)$ and cavity coupling $(\Omega/\omega_c)$. The spectrum is superimposed with the spectral representation of 5 fs (shaded in orange) and 400 fs (shaded in yellow, purple, cyan, red and blue, from left to right) RCPs with resonant frequencies at, above and below the CI
     (marked \textcolor{magenta}{$\times$} on the energy axis).
     (b, c) Time-dependent cavity polarization ($\tilde{P}$=$\mathcal{P}_{k_{-1}}(t)/\hbar/p_{ex}(t_f)$) generated by the pulses shown in (a), respectively. Here the line(s) color follows that of the excitation pulse(s) in (a).
     The time-dependent net polarizations are
     normalized by the total excitation probability $p_{ex}(t_f)$ after the
     pulse (cf. main text).} 
     %
    % \\\textcolor{red}{[Please change the labeling of the Figures,
    % c$\to$b, d$\to$c, f$\to$d. Please change legend to $\kappa/\omega$.]}
    %}
  %The $\kappa=0.032$ a.u.,  implies $\frac{\kappa}{\omega}\approx$20 for $\omega$=0.007 a.u.}
 \end{figure}

\emph{Interaction of the cavity Jahn-Teller system with circularly polarized light --} %
We now couple the CJT system to an external circularly polarized (CP) pulse propagating along the $z$-direction, and
consider a stronger JT coupling, $\kappa/\omega\approx 8.8$, than before. In this strong coupling regime, the
electronic states are highly mixed, which in turn contributes to a strong polarization mixing in the cavity.
We consider
a cavity coupling of $\Omega/\omega_c$ $\approx 0.1$.
Depending on the bandwidth of the CP pulse, it can be made resonant with a
single $|k_j\rangle$ state, or a group of states.
The interaction of the CP pulse with the CJT system is treated in the electric dipole approximation,
\mbox{$ - (\hat{\mu}_x \mathcal{E}_x(t) + \hat{\mu}_y \mathcal{E}_y(t))$}.
The transition dipole operators $\hat{\mu}_x$ and $\hat{\mu}_y$
are assumed to couple the cavity ground state with the one-photon
states of either $x$ or $y$ polarization, respectively. The left and right CP pulses, namely LCP and RCP, are
taken as Fourier-limited of Gaussian shape. Details are found in Ref.~\cite{nandipati2021dynamical}. 
Dipolar transitions initiated from the GS $|A,0_+,0_-\rangle$ fulfill the selection rule
$\Delta j=\pm1$ and CP radiation can only induce transitions
$(0 \leftrightarrow 1)$ for LCP and $(0 \leftrightarrow -1)$ for
RCP~\cite{nandipati2020generation}.
We consider the CP pulses weak enough to remain in first-order perturbation so that the generated polarization is intensity-independent when normalized by the excited-state population $p_{ex}(t_f)=(1-p_0(t_f))$, where $p_0(t)$ is the population of the absolute ground state and $t_f$ is the final time of the simulation after the pulse is over. 
  %$p_{ex}(t_f)$ is kept in the order of 5$\times 10^{-2}$ or lower.
  %
  
  We apply both long (400~fs, shaded in yellow/purple/cyan/red/blue) and short (5~fs, shaded in orange) RCP pulses whose spectra are shown superimposed in Figure~\ref{el_ph_Imbalance_gen}a.
  At about 7.4 and 7.52~eV, the long pulses (red/blue) are resonant with the eigenstates featuring successively the largest transition dipole moments in the UP branch, located above the CI.
  The bandwidths of the pulses are narrow enough to target only these eigenstates, which result in
  a stationary polarization after their action is over, as indicated by the red and blue curves in Figure~\ref{el_ph_Imbalance_gen}b.
  The most optically bright state presents the net polarization $\tilde{\mathcal{P}}_{k_{-1}}$ about 0.009~$\hbar$ in clockwise direction (cf. solid red curve) which is in opposite direction to the polarization presented by the second-most optically bright state of about -0.022~$\hbar$ (cf. solid blue curve). 
  %On the other hand, the ring currents in these two states possess same direction with different magnitudes (cf. dashed red/blue curves).
  %
  The long pulses (yellow/purple/cyan) targeting the states in the LP branch, located below the CI,
  result now in polarizations different polarization directions and with different magnitudes,
  as shown in Figure~\ref{el_ph_Imbalance_gen}b, in the same colors as that of the corresponding pulses.
  Thus, different spectral regions of the CJT system feature
  opposite cavity-polarization directions
  and these states can be addressed by pulses of fixed polarization (e.g. RCP) with different frequencies.
  This is a direct consequence of the vibronic coupling effects and of the mixing of the photonic
  and vibronic angular momenta.
  Therefore, an RCP pulse does not necessarily result in $(-)$ polarization of the cavity,
  as is always the case in the $\kappa\to0$ limit. 

 Both the LP and UP branches together can be spanned by a pulse as short as 5~fs centered at 7~eV. 
 Since the eigenstates inside the pulse bandwidth have either $(+)$ or $(-)$ net expectation of the photonic
 angular momentum, the resulting polarization is highly oscillatory (cf. Figure~\ref{el_ph_Imbalance_gen}c).
 The oscillatory circular polarization dynamics of the CJT system reaches polarization amplitudes
 markedly larger than the maximum polarization achieved by long pulses
 targeting single eigenstates.
 %In addition to the overall slower oscillation dynamics owing to the eigenstate(s) average energy gap, the short pulse triggers faster Rabi-oscillations (time period ~5 fs) owing to the polaritonic gap ($\Omega$) between the LP and UP branches. %This hints at the possibility to control polaritonic phenomena on attosecond timescales.
  %Also, the magnitudes of time-averaged currents are quite small as compared to that of 10 fs pulse above.
  %
  
\emph{Conclusions --} 
 In summary, we have described the mechanism of formation of mixed-polarization
 JT polaritonic states by the interaction of JT-active systems
with the two degenerate cavity modes at normal incidence of a FP resonator.
%investigated the properties of these states in terms of their cavity polarization content. 
  %
  The upper and lower JT polaritons present CIs that strongly mix the two
  polarization directions of the cavity photon.
  This vibronic coupling suppresses the maximum degree of photonic angular momentum
  of individual eigenstates of the polaritonic system. However, short circularly polarized
  external pulses can trigger highly-oscillatory polarization dynamics as a superposition of the
  polaritonic eigenstates.
  These findings can result in schemes to achieve chiral environments in cavities without
  resorting to special types of mirrors. The ultrafast polarization oscillations
  resulting from the vibronic-photonic coupling
  can be verified by pump-probe spectroscopic measurements involving CP pulses~\cite{svoboda2022generation,baykusheva2016bicircular}.
  %and the amount of ring currents presented by the JTP states.     
  %

 \begin{acknowledgments}
    This work has been supported by the collaborative research center ``SFB 1249: N-Heteropolyzyklen als Funktionsmaterialen''
    of the German Research Foundation (DFG). We thank professor Wolfgang Domcke for useful comments on the manuscript.
\end{acknowledgments}

%\pagebreak
%\bibliography{JT-cavity}

%merlin.mbs apsrev4-1.bst 2010-07-25 4.21a (PWD, AO, DPC) hacked
%Control: key (0)
%Control: author (8) initials jnrlst
%Control: editor formatted (1) identically to author
%Control: production of article title (-1) disabled
%Control: page (0) single
%Control: year (1) truncated
%Control: production of eprint (0) enabled
\begin{thebibliography}{39}%
\makeatletter
\providecommand \@ifxundefined [1]{%
 \@ifx{#1\undefined}
}%
\providecommand \@ifnum [1]{%
 \ifnum #1\expandafter \@firstoftwo
 \else \expandafter \@secondoftwo
 \fi
}%
\providecommand \@ifx [1]{%
 \ifx #1\expandafter \@firstoftwo
 \else \expandafter \@secondoftwo
 \fi
}%
\providecommand \natexlab [1]{#1}%
\providecommand \enquote  [1]{``#1''}%
\providecommand \bibnamefont  [1]{#1}%
\providecommand \bibfnamefont [1]{#1}%
\providecommand \citenamefont [1]{#1}%
\providecommand \href@noop [0]{\@secondoftwo}%
\providecommand \href [0]{\begingroup \@sanitize@url \@href}%
\providecommand \@href[1]{\@@startlink{#1}\@@href}%
\providecommand \@@href[1]{\endgroup#1\@@endlink}%
\providecommand \@sanitize@url [0]{\catcode `\\12\catcode `\$12\catcode
  `\&12\catcode `\#12\catcode `\^12\catcode `\_12\catcode `\%12\relax}%
\providecommand \@@startlink[1]{}%
\providecommand \@@endlink[0]{}%
\providecommand \url  [0]{\begingroup\@sanitize@url \@url }%
\providecommand \@url [1]{\endgroup\@href {#1}{\urlprefix }}%
\providecommand \urlprefix  [0]{URL }%
\providecommand \Eprint [0]{\href }%
\providecommand \doibase [0]{http://dx.doi.org/}%
\providecommand \selectlanguage [0]{\@gobble}%
\providecommand \bibinfo  [0]{\@secondoftwo}%
\providecommand \bibfield  [0]{\@secondoftwo}%
\providecommand \translation [1]{[#1]}%
\providecommand \BibitemOpen [0]{}%
\providecommand \bibitemStop [0]{}%
\providecommand \bibitemNoStop [0]{.\EOS\space}%
\providecommand \EOS [0]{\spacefactor3000\relax}%
\providecommand \BibitemShut  [1]{\csname bibitem#1\endcsname}%
\let\auto@bib@innerbib\@empty
%</preamble>
\bibitem [{\citenamefont {Hutchison}\ \emph {et~al.}(2012)\citenamefont
  {Hutchison}, \citenamefont {Schwartz}, \citenamefont {Genet}, \citenamefont
  {Devaux},\ and\ \citenamefont {Ebbesen}}]{hutchison2012modifying}%
  \BibitemOpen
  \bibfield  {author} {\bibinfo {author} {\bibfnamefont {J.~A.}\ \bibnamefont
  {Hutchison}}, \bibinfo {author} {\bibfnamefont {T.}~\bibnamefont {Schwartz}},
  \bibinfo {author} {\bibfnamefont {C.}~\bibnamefont {Genet}}, \bibinfo
  {author} {\bibfnamefont {E.}~\bibnamefont {Devaux}}, \ and\ \bibinfo {author}
  {\bibfnamefont {T.~W.}\ \bibnamefont {Ebbesen}},\ }\href@noop {} {\bibfield
  {journal} {\bibinfo  {journal} {Angew. Chem.}\ }\textbf {\bibinfo {volume}
  {51}},\ \bibinfo {pages} {1592} (\bibinfo {year} {2012})}\BibitemShut
  {NoStop}%
\bibitem [{\citenamefont {Schwartz}\ \emph {et~al.}(2013)\citenamefont
  {Schwartz}, \citenamefont {Hutchison}, \citenamefont {L{\'e}onard},
  \citenamefont {Genet}, \citenamefont {Haacke},\ and\ \citenamefont
  {Ebbesen}}]{schwartz2013polariton}%
  \BibitemOpen
  \bibfield  {author} {\bibinfo {author} {\bibfnamefont {T.}~\bibnamefont
  {Schwartz}}, \bibinfo {author} {\bibfnamefont {J.~A.}\ \bibnamefont
  {Hutchison}}, \bibinfo {author} {\bibfnamefont {J.}~\bibnamefont
  {L{\'e}onard}}, \bibinfo {author} {\bibfnamefont {C.}~\bibnamefont {Genet}},
  \bibinfo {author} {\bibfnamefont {S.}~\bibnamefont {Haacke}}, \ and\ \bibinfo
  {author} {\bibfnamefont {T.~W.}\ \bibnamefont {Ebbesen}},\ }\href@noop {}
  {\bibfield  {journal} {\bibinfo  {journal} {ChemPhysChem}\ }\textbf {\bibinfo
  {volume} {14}},\ \bibinfo {pages} {125} (\bibinfo {year} {2013})}\BibitemShut
  {NoStop}%
\bibitem [{\citenamefont {Herrera}\ and\ \citenamefont
  {Owrutsky}(2020)}]{herrera2020molecular}%
  \BibitemOpen
  \bibfield  {author} {\bibinfo {author} {\bibfnamefont {F.}~\bibnamefont
  {Herrera}}\ and\ \bibinfo {author} {\bibfnamefont {J.}~\bibnamefont
  {Owrutsky}},\ }\href@noop {} {\bibfield  {journal} {\bibinfo  {journal} {J.
  Chem. Phys.}\ }\textbf {\bibinfo {volume} {152}},\ \bibinfo {pages} {100902}
  (\bibinfo {year} {2020})}\BibitemShut {NoStop}%
\bibitem [{\citenamefont {Ribeiro}\ \emph {et~al.}(2018)\citenamefont
  {Ribeiro}, \citenamefont {Mart{\'\i}nez-Mart{\'\i}nez}, \citenamefont {Du},
  \citenamefont {Campos-Gonzalez-Angulo},\ and\ \citenamefont
  {Yuen-Zhou}}]{ribeiro2018polariton}%
  \BibitemOpen
  \bibfield  {author} {\bibinfo {author} {\bibfnamefont {R.~F.}\ \bibnamefont
  {Ribeiro}}, \bibinfo {author} {\bibfnamefont {L.~A.}\ \bibnamefont
  {Mart{\'\i}nez-Mart{\'\i}nez}}, \bibinfo {author} {\bibfnamefont
  {M.}~\bibnamefont {Du}}, \bibinfo {author} {\bibfnamefont {J.}~\bibnamefont
  {Campos-Gonzalez-Angulo}}, \ and\ \bibinfo {author} {\bibfnamefont
  {J.}~\bibnamefont {Yuen-Zhou}},\ }\href@noop {} {\bibfield  {journal}
  {\bibinfo  {journal} {Chem. Sci.}\ }\textbf {\bibinfo {volume} {9}},\
  \bibinfo {pages} {6325} (\bibinfo {year} {2018})}\BibitemShut {NoStop}%
\bibitem [{\citenamefont {Kowalewski}\ \emph {et~al.}(2016)\citenamefont
  {Kowalewski}, \citenamefont {Bennett},\ and\ \citenamefont
  {Mukamel}}]{kowalewski2016non}%
  \BibitemOpen
  \bibfield  {author} {\bibinfo {author} {\bibfnamefont {M.}~\bibnamefont
  {Kowalewski}}, \bibinfo {author} {\bibfnamefont {K.}~\bibnamefont {Bennett}},
  \ and\ \bibinfo {author} {\bibfnamefont {S.}~\bibnamefont {Mukamel}},\
  }\href@noop {} {\bibfield  {journal} {\bibinfo  {journal} {J. Chem. Phys.}\
  }\textbf {\bibinfo {volume} {144}},\ \bibinfo {pages} {054309} (\bibinfo
  {year} {2016})}\BibitemShut {NoStop}%
\bibitem [{\citenamefont {Galego}\ \emph {et~al.}(2016)\citenamefont {Galego},
  \citenamefont {Garcia-Vidal},\ and\ \citenamefont
  {Feist}}]{galego2016suppressing}%
  \BibitemOpen
  \bibfield  {author} {\bibinfo {author} {\bibfnamefont {J.}~\bibnamefont
  {Galego}}, \bibinfo {author} {\bibfnamefont {F.~J.}\ \bibnamefont
  {Garcia-Vidal}}, \ and\ \bibinfo {author} {\bibfnamefont {J.}~\bibnamefont
  {Feist}},\ }\href@noop {} {\bibfield  {journal} {\bibinfo  {journal} {Nat.
  Comm.}\ }\textbf {\bibinfo {volume} {7}},\ \bibinfo {pages} {1} (\bibinfo
  {year} {2016})}\BibitemShut {NoStop}%
\bibitem [{\citenamefont {Flick}\ \emph {et~al.}(2017)\citenamefont {Flick},
  \citenamefont {Ruggenthaler}, \citenamefont {Appel},\ and\ \citenamefont
  {Rubio}}]{flick2017atoms}%
  \BibitemOpen
  \bibfield  {author} {\bibinfo {author} {\bibfnamefont {J.}~\bibnamefont
  {Flick}}, \bibinfo {author} {\bibfnamefont {M.}~\bibnamefont {Ruggenthaler}},
  \bibinfo {author} {\bibfnamefont {H.}~\bibnamefont {Appel}}, \ and\ \bibinfo
  {author} {\bibfnamefont {A.}~\bibnamefont {Rubio}},\ }\href@noop {}
  {\bibfield  {journal} {\bibinfo  {journal} {Proc. Nat. Acad. Sci.}\ }\textbf
  {\bibinfo {volume} {114}},\ \bibinfo {pages} {3026} (\bibinfo {year}
  {2017})}\BibitemShut {NoStop}%
\bibitem [{\citenamefont {Feist}\ \emph {et~al.}(2018)\citenamefont {Feist},
  \citenamefont {Galego},\ and\ \citenamefont
  {Garcia-Vidal}}]{feist2018polaritonic}%
  \BibitemOpen
  \bibfield  {author} {\bibinfo {author} {\bibfnamefont {J.}~\bibnamefont
  {Feist}}, \bibinfo {author} {\bibfnamefont {J.}~\bibnamefont {Galego}}, \
  and\ \bibinfo {author} {\bibfnamefont {F.~J.}\ \bibnamefont {Garcia-Vidal}},\
  }\href@noop {} {\bibfield  {journal} {\bibinfo  {journal} {ACS Photonics}\
  }\textbf {\bibinfo {volume} {5}},\ \bibinfo {pages} {205} (\bibinfo {year}
  {2018})}\BibitemShut {NoStop}%
\bibitem [{\citenamefont
  {Vendrell}(2018{\natexlab{a}})}]{vendrell2018coherent}%
  \BibitemOpen
  \bibfield  {author} {\bibinfo {author} {\bibfnamefont {O.}~\bibnamefont
  {Vendrell}},\ }\href@noop {} {\bibfield  {journal} {\bibinfo  {journal}
  {Chem. Phys.}\ }\textbf {\bibinfo {volume} {509}},\ \bibinfo {pages} {55}
  (\bibinfo {year} {2018}{\natexlab{a}})}\BibitemShut {NoStop}%
\bibitem [{\citenamefont {Herrera}\ and\ \citenamefont
  {Spano}(2016)}]{herrera2016cavity}%
  \BibitemOpen
  \bibfield  {author} {\bibinfo {author} {\bibfnamefont {F.}~\bibnamefont
  {Herrera}}\ and\ \bibinfo {author} {\bibfnamefont {F.~C.}\ \bibnamefont
  {Spano}},\ }\href@noop {} {\bibfield  {journal} {\bibinfo  {journal} {Phys.
  Rev. Lett.}\ }\textbf {\bibinfo {volume} {116}},\ \bibinfo {pages} {238301}
  (\bibinfo {year} {2016})}\BibitemShut {NoStop}%
\bibitem [{\citenamefont {Herrera}\ and\ \citenamefont
  {Spano}(2017)}]{herrera2017dark}%
  \BibitemOpen
  \bibfield  {author} {\bibinfo {author} {\bibfnamefont {F.}~\bibnamefont
  {Herrera}}\ and\ \bibinfo {author} {\bibfnamefont {F.~C.}\ \bibnamefont
  {Spano}},\ }\href@noop {} {\bibfield  {journal} {\bibinfo  {journal} {Phys.
  Rev. Lett.}\ }\textbf {\bibinfo {volume} {118}},\ \bibinfo {pages} {223601}
  (\bibinfo {year} {2017})}\BibitemShut {NoStop}%
\bibitem [{\citenamefont
  {Vendrell}(2018{\natexlab{b}})}]{vendrell2018collective}%
  \BibitemOpen
  \bibfield  {author} {\bibinfo {author} {\bibfnamefont {O.}~\bibnamefont
  {Vendrell}},\ }\href@noop {} {\bibfield  {journal} {\bibinfo  {journal}
  {Phys. Rev. Lett.}\ }\textbf {\bibinfo {volume} {121}},\ \bibinfo {pages}
  {253001} (\bibinfo {year} {2018}{\natexlab{b}})}\BibitemShut {NoStop}%
\bibitem [{\citenamefont {Dunkelberger}\ \emph {et~al.}(2022)\citenamefont
  {Dunkelberger}, \citenamefont {Simpkins}, \citenamefont {Vurgaftman},\ and\
  \citenamefont {Owrutsky}}]{dunkelberger2022vibration}%
  \BibitemOpen
  \bibfield  {author} {\bibinfo {author} {\bibfnamefont {A.~D.}\ \bibnamefont
  {Dunkelberger}}, \bibinfo {author} {\bibfnamefont {B.~S.}\ \bibnamefont
  {Simpkins}}, \bibinfo {author} {\bibfnamefont {I.}~\bibnamefont
  {Vurgaftman}}, \ and\ \bibinfo {author} {\bibfnamefont {J.~C.}\ \bibnamefont
  {Owrutsky}},\ }\href@noop {} {\bibfield  {journal} {\bibinfo  {journal} {Ann.
  Rev. Phys. Chem.}\ }\textbf {\bibinfo {volume} {73}} (\bibinfo {year}
  {2022})}\BibitemShut {NoStop}%
\bibitem [{\citenamefont {Morigi}\ \emph {et~al.}(2007)\citenamefont {Morigi},
  \citenamefont {Pinkse}, \citenamefont {Kowalewski},\ and\ \citenamefont
  {de~Vivie-Riedle}}]{morigi2007cavity}%
  \BibitemOpen
  \bibfield  {author} {\bibinfo {author} {\bibfnamefont {G.}~\bibnamefont
  {Morigi}}, \bibinfo {author} {\bibfnamefont {P.~W.}\ \bibnamefont {Pinkse}},
  \bibinfo {author} {\bibfnamefont {M.}~\bibnamefont {Kowalewski}}, \ and\
  \bibinfo {author} {\bibfnamefont {R.}~\bibnamefont {de~Vivie-Riedle}},\
  }\href@noop {} {\bibfield  {journal} {\bibinfo  {journal} {Phys. Rev. Lett.}\
  }\textbf {\bibinfo {volume} {99}},\ \bibinfo {pages} {073001} (\bibinfo
  {year} {2007})}\BibitemShut {NoStop}%
\bibitem [{\citenamefont {Ebbesen}(2016)}]{ebbesen2016hybrid}%
  \BibitemOpen
  \bibfield  {author} {\bibinfo {author} {\bibfnamefont {T.~W.}\ \bibnamefont
  {Ebbesen}},\ }\href@noop {} {\bibfield  {journal} {\bibinfo  {journal} {Acc.
  Chem. Res.}\ }\textbf {\bibinfo {volume} {49}},\ \bibinfo {pages} {2403}
  (\bibinfo {year} {2016})}\BibitemShut {NoStop}%
\bibitem [{\citenamefont {Orgiu}\ \emph {et~al.}(2015)\citenamefont {Orgiu},
  \citenamefont {George}, \citenamefont {Hutchison}, \citenamefont {Devaux},
  \citenamefont {Dayen}, \citenamefont {Doudin}, \citenamefont {Stellacci},
  \citenamefont {Genet}, \citenamefont {Schachenmayer}, \citenamefont {Genes}
  \emph {et~al.}}]{orgiu2015conductivity}%
  \BibitemOpen
  \bibfield  {author} {\bibinfo {author} {\bibfnamefont {E.}~\bibnamefont
  {Orgiu}}, \bibinfo {author} {\bibfnamefont {J.}~\bibnamefont {George}},
  \bibinfo {author} {\bibfnamefont {J.}~\bibnamefont {Hutchison}}, \bibinfo
  {author} {\bibfnamefont {E.}~\bibnamefont {Devaux}}, \bibinfo {author}
  {\bibfnamefont {J.}~\bibnamefont {Dayen}}, \bibinfo {author} {\bibfnamefont
  {B.}~\bibnamefont {Doudin}}, \bibinfo {author} {\bibfnamefont
  {F.}~\bibnamefont {Stellacci}}, \bibinfo {author} {\bibfnamefont
  {C.}~\bibnamefont {Genet}}, \bibinfo {author} {\bibfnamefont
  {J.}~\bibnamefont {Schachenmayer}}, \bibinfo {author} {\bibfnamefont
  {C.}~\bibnamefont {Genes}},  \emph {et~al.},\ }\href@noop {} {\bibfield
  {journal} {\bibinfo  {journal} {Nat. Mater.}\ }\textbf {\bibinfo {volume}
  {14}},\ \bibinfo {pages} {1123} (\bibinfo {year} {2015})}\BibitemShut
  {NoStop}%
\bibitem [{\citenamefont {Wang}\ \emph {et~al.}(2019)\citenamefont {Wang},
  \citenamefont {Ronca},\ and\ \citenamefont {Sentef}}]{wang2019cavity}%
  \BibitemOpen
  \bibfield  {author} {\bibinfo {author} {\bibfnamefont {X.}~\bibnamefont
  {Wang}}, \bibinfo {author} {\bibfnamefont {E.}~\bibnamefont {Ronca}}, \ and\
  \bibinfo {author} {\bibfnamefont {M.~A.}\ \bibnamefont {Sentef}},\
  }\href@noop {} {\bibfield  {journal} {\bibinfo  {journal} {Phys. Rev. B}\
  }\textbf {\bibinfo {volume} {99}},\ \bibinfo {pages} {235156} (\bibinfo
  {year} {2019})}\BibitemShut {NoStop}%
\bibitem [{\citenamefont {Gautier}\ \emph {et~al.}(2022)\citenamefont
  {Gautier}, \citenamefont {Li}, \citenamefont {Ebbesen},\ and\ \citenamefont
  {Genet}}]{gautier2022planar}%
  \BibitemOpen
  \bibfield  {author} {\bibinfo {author} {\bibfnamefont {J.}~\bibnamefont
  {Gautier}}, \bibinfo {author} {\bibfnamefont {M.}~\bibnamefont {Li}},
  \bibinfo {author} {\bibfnamefont {T.~W.}\ \bibnamefont {Ebbesen}}, \ and\
  \bibinfo {author} {\bibfnamefont {C.}~\bibnamefont {Genet}},\ }\href@noop {}
  {\bibfield  {journal} {\bibinfo  {journal} {ACS photonics}\ }\textbf
  {\bibinfo {volume} {9}},\ \bibinfo {pages} {778} (\bibinfo {year}
  {2022})}\BibitemShut {NoStop}%
\bibitem [{\citenamefont {H{\"u}bener}\ \emph {et~al.}(2021)\citenamefont
  {H{\"u}bener}, \citenamefont {De~Giovannini}, \citenamefont {Sch{\"a}fer},
  \citenamefont {Andberger}, \citenamefont {Ruggenthaler}, \citenamefont
  {Faist},\ and\ \citenamefont {Rubio}}]{hubener2021engineering}%
  \BibitemOpen
  \bibfield  {author} {\bibinfo {author} {\bibfnamefont {H.}~\bibnamefont
  {H{\"u}bener}}, \bibinfo {author} {\bibfnamefont {U.}~\bibnamefont
  {De~Giovannini}}, \bibinfo {author} {\bibfnamefont {C.}~\bibnamefont
  {Sch{\"a}fer}}, \bibinfo {author} {\bibfnamefont {J.}~\bibnamefont
  {Andberger}}, \bibinfo {author} {\bibfnamefont {M.}~\bibnamefont
  {Ruggenthaler}}, \bibinfo {author} {\bibfnamefont {J.}~\bibnamefont {Faist}},
  \ and\ \bibinfo {author} {\bibfnamefont {A.}~\bibnamefont {Rubio}},\
  }\href@noop {} {\bibfield  {journal} {\bibinfo  {journal} {Nat. Mater.}\
  }\textbf {\bibinfo {volume} {20}},\ \bibinfo {pages} {438} (\bibinfo {year}
  {2021})}\BibitemShut {NoStop}%
\bibitem [{\citenamefont {Feis}\ \emph {et~al.}(2020)\citenamefont {Feis},
  \citenamefont {Beutel}, \citenamefont {K{\"o}pfler}, \citenamefont
  {Garcia-Santiago}, \citenamefont {Rockstuhl}, \citenamefont {Wegener},\ and\
  \citenamefont {Fernandez-Corbaton}}]{feis2020helicity}%
  \BibitemOpen
  \bibfield  {author} {\bibinfo {author} {\bibfnamefont {J.}~\bibnamefont
  {Feis}}, \bibinfo {author} {\bibfnamefont {D.}~\bibnamefont {Beutel}},
  \bibinfo {author} {\bibfnamefont {J.}~\bibnamefont {K{\"o}pfler}}, \bibinfo
  {author} {\bibfnamefont {X.}~\bibnamefont {Garcia-Santiago}}, \bibinfo
  {author} {\bibfnamefont {C.}~\bibnamefont {Rockstuhl}}, \bibinfo {author}
  {\bibfnamefont {M.}~\bibnamefont {Wegener}}, \ and\ \bibinfo {author}
  {\bibfnamefont {I.}~\bibnamefont {Fernandez-Corbaton}},\ }\href@noop {}
  {\bibfield  {journal} {\bibinfo  {journal} {Phys. Rev. Lett.}\ }\textbf
  {\bibinfo {volume} {124}},\ \bibinfo {pages} {033201} (\bibinfo {year}
  {2020})}\BibitemShut {NoStop}%
\bibitem [{\citenamefont {Shelykh}\ \emph {et~al.}(2009)\citenamefont
  {Shelykh}, \citenamefont {Kavokin}, \citenamefont {Rubo}, \citenamefont
  {Liew},\ and\ \citenamefont {Malpuech}}]{shelykh2009polariton}%
  \BibitemOpen
  \bibfield  {author} {\bibinfo {author} {\bibfnamefont {I.~A.}\ \bibnamefont
  {Shelykh}}, \bibinfo {author} {\bibfnamefont {A.~V.}\ \bibnamefont
  {Kavokin}}, \bibinfo {author} {\bibfnamefont {Y.~G.}\ \bibnamefont {Rubo}},
  \bibinfo {author} {\bibfnamefont {T.}~\bibnamefont {Liew}}, \ and\ \bibinfo
  {author} {\bibfnamefont {G.}~\bibnamefont {Malpuech}},\ }\href@noop {}
  {\bibfield  {journal} {\bibinfo  {journal} {Semicond. Sci. Technol.}\
  }\textbf {\bibinfo {volume} {25}},\ \bibinfo {pages} {013001} (\bibinfo
  {year} {2009})}\BibitemShut {NoStop}%
\bibitem [{\citenamefont {Yoo}\ and\ \citenamefont
  {Park}(2015)}]{yoo2015chiral}%
  \BibitemOpen
  \bibfield  {author} {\bibinfo {author} {\bibfnamefont {S.}~\bibnamefont
  {Yoo}}\ and\ \bibinfo {author} {\bibfnamefont {Q.-H.}\ \bibnamefont {Park}},\
  }\href@noop {} {\bibfield  {journal} {\bibinfo  {journal} {Phys. Rev. Lett.}\
  }\textbf {\bibinfo {volume} {114}},\ \bibinfo {pages} {203003} (\bibinfo
  {year} {2015})}\BibitemShut {NoStop}%
\bibitem [{\citenamefont {Sun}\ \emph {et~al.}(2022)\citenamefont {Sun},
  \citenamefont {Gu},\ and\ \citenamefont {Mukamel}}]{sun2022polariton}%
  \BibitemOpen
  \bibfield  {author} {\bibinfo {author} {\bibfnamefont {S.}~\bibnamefont
  {Sun}}, \bibinfo {author} {\bibfnamefont {B.}~\bibnamefont {Gu}}, \ and\
  \bibinfo {author} {\bibfnamefont {S.}~\bibnamefont {Mukamel}},\ }\href@noop
  {} {\bibfield  {journal} {\bibinfo  {journal} {Chem. Sci.}\ } (\bibinfo
  {year} {2022})}\BibitemShut {NoStop}%
\bibitem [{\citenamefont {Abasahl}\ \emph {et~al.}(2013)\citenamefont
  {Abasahl}, \citenamefont {Dutta-Gupta}, \citenamefont {Santschi},\ and\
  \citenamefont {Martin}}]{abasahl2013coupling}%
  \BibitemOpen
  \bibfield  {author} {\bibinfo {author} {\bibfnamefont {B.}~\bibnamefont
  {Abasahl}}, \bibinfo {author} {\bibfnamefont {S.}~\bibnamefont
  {Dutta-Gupta}}, \bibinfo {author} {\bibfnamefont {C.}~\bibnamefont
  {Santschi}}, \ and\ \bibinfo {author} {\bibfnamefont {O.~J.}\ \bibnamefont
  {Martin}},\ }\href@noop {} {\bibfield  {journal} {\bibinfo  {journal} {Nano
  Lett.}\ }\textbf {\bibinfo {volume} {13}},\ \bibinfo {pages} {4575} (\bibinfo
  {year} {2013})}\BibitemShut {NoStop}%
\bibitem [{\citenamefont {Bersuker}(2006)}]{bersuker2006jahn}%
  \BibitemOpen
  \bibfield  {author} {\bibinfo {author} {\bibfnamefont {I.~B.}\ \bibnamefont
  {Bersuker}},\ }\href@noop {} {\emph {\bibinfo {title} {The Jahn-Teller
  effect}}}\ (\bibinfo  {publisher} {Cambridge University Press},\ \bibinfo
  {year} {2006})\BibitemShut {NoStop}%
\bibitem [{\citenamefont {Englman}(1972)}]{englman}%
  \BibitemOpen
  \bibfield  {author} {\bibinfo {author} {\bibfnamefont {R.}~\bibnamefont
  {Englman}},\ }\href@noop {} {\emph {\bibinfo {title} {The {Jahn--Teller}
  Effect in Molecules and Crystals}}}\ (\bibinfo  {publisher} {Wiley, New
  York},\ \bibinfo {year} {1972})\BibitemShut {NoStop}%
\bibitem [{\citenamefont {Longuet-Higgins}\ \emph {et~al.}(1958)\citenamefont
  {Longuet-Higgins}, \citenamefont {{\"{O}}pik}, \citenamefont {Pryce},\ and\
  \citenamefont {Sack}}]{Longuet-Higgins}%
  \BibitemOpen
  \bibfield  {author} {\bibinfo {author} {\bibfnamefont {H.~C.}\ \bibnamefont
  {Longuet-Higgins}}, \bibinfo {author} {\bibfnamefont {U.}~\bibnamefont
  {{\"{O}}pik}}, \bibinfo {author} {\bibfnamefont {M.~H.~L.}\ \bibnamefont
  {Pryce}}, \ and\ \bibinfo {author} {\bibfnamefont {R.~A.}\ \bibnamefont
  {Sack}},\ }\href {\doibase 10.1098/rspa.1958.0022} {\bibfield  {journal}
  {\bibinfo  {journal} {Proc. R. Soc. A}\ }\textbf {\bibinfo {volume} {244}},\
  \bibinfo {pages} {1} (\bibinfo {year} {1958})}\BibitemShut {NoStop}%
\bibitem [{\citenamefont {K{\"{o}}ppel}\ \emph {et~al.}(1984)\citenamefont
  {K{\"{o}}ppel}, \citenamefont {Domcke},\ and\ \citenamefont
  {Cederbaum}}]{koppel}%
  \BibitemOpen
  \bibfield  {author} {\bibinfo {author} {\bibfnamefont {H.}~\bibnamefont
  {K{\"{o}}ppel}}, \bibinfo {author} {\bibfnamefont {W.}~\bibnamefont
  {Domcke}}, \ and\ \bibinfo {author} {\bibfnamefont {L.}~\bibnamefont
  {Cederbaum}},\ }\href {\doibase 10.1002/9780470142813.ch2} {\bibfield
  {journal} {\bibinfo  {journal} {Adv. Chem. Phys.}\ }\textbf {\bibinfo
  {volume} {57}},\ \bibinfo {pages} {59} (\bibinfo {year} {1984})}\BibitemShut
  {NoStop}%
\bibitem [{\citenamefont {Whetten}\ \emph {et~al.}(1986)\citenamefont
  {Whetten}, \citenamefont {Haber},\ and\ \citenamefont
  {Grant}}]{whetten1986dynamic}%
  \BibitemOpen
  \bibfield  {author} {\bibinfo {author} {\bibfnamefont {R.~L.}\ \bibnamefont
  {Whetten}}, \bibinfo {author} {\bibfnamefont {K.~S.}\ \bibnamefont {Haber}},
  \ and\ \bibinfo {author} {\bibfnamefont {E.~R.}\ \bibnamefont {Grant}},\
  }\href@noop {} {\bibfield  {journal} {\bibinfo  {journal} {J. Chem. Phys.}\
  }\textbf {\bibinfo {volume} {84}},\ \bibinfo {pages} {1270} (\bibinfo {year}
  {1986})}\BibitemShut {NoStop}%
\bibitem [{\citenamefont {Jahn}\ and\ \citenamefont
  {Teller}(1937)}]{jahn1937stability}%
  \BibitemOpen
  \bibfield  {author} {\bibinfo {author} {\bibfnamefont {H.~A.}\ \bibnamefont
  {Jahn}}\ and\ \bibinfo {author} {\bibfnamefont {E.}~\bibnamefont {Teller}},\
  }\href@noop {} {\bibfield  {journal} {\bibinfo  {journal} {Proc. R. Soc. A}\
  }\textbf {\bibinfo {volume} {161}},\ \bibinfo {pages} {220} (\bibinfo {year}
  {1937})}\BibitemShut {NoStop}%
\bibitem [{\citenamefont {Moffitt}\ and\ \citenamefont
  {Liehr}(1957)}]{moffitt1957configurational}%
  \BibitemOpen
  \bibfield  {author} {\bibinfo {author} {\bibfnamefont {W.}~\bibnamefont
  {Moffitt}}\ and\ \bibinfo {author} {\bibfnamefont {A.}~\bibnamefont
  {Liehr}},\ }\href@noop {} {\bibfield  {journal} {\bibinfo  {journal} {Phys.
  Rev.}\ }\textbf {\bibinfo {volume} {106}},\ \bibinfo {pages} {1195} (\bibinfo
  {year} {1957})}\BibitemShut {NoStop}%
\bibitem [{\citenamefont {Sturge}(1968)}]{sturge}%
  \BibitemOpen
  \bibfield  {author} {\bibinfo {author} {\bibfnamefont {M.~D.}\ \bibnamefont
  {Sturge}},\ }\href {\doibase 10.1016/S0081-1947(08)60218-0} {\bibfield
  {journal} {\bibinfo  {journal} {Solid St. Phys.}\ }\textbf {\bibinfo {volume}
  {20}},\ \bibinfo {pages} {91} (\bibinfo {year} {1968})}\BibitemShut {NoStop}%
\bibitem [{\citenamefont {Tannor}(2007)}]{tannor2007introduction}%
  \BibitemOpen
  \bibfield  {author} {\bibinfo {author} {\bibfnamefont {D.~J.}\ \bibnamefont
  {Tannor}},\ }\href@noop {} {\emph {\bibinfo {title} {Introduction to quantum
  mechanics: a time-dependent perspective}}}\ (\bibinfo  {publisher}
  {University Science Books},\ \bibinfo {year} {2007})\BibitemShut {NoStop}%
\bibitem [{\citenamefont {Crisp}(1991)}]{cri_91_2430}%
  \BibitemOpen
  \bibfield  {author} {\bibinfo {author} {\bibfnamefont {M.~D.}\ \bibnamefont
  {Crisp}},\ }\href {\doibase 10.1103/PhysRevA.43.2430} {\bibfield  {journal}
  {\bibinfo  {journal} {Phys. Rev. A}\ }\textbf {\bibinfo {volume} {43}},\
  \bibinfo {pages} {2430} (\bibinfo {year} {1991})}\BibitemShut {NoStop}%
\bibitem [{\citenamefont {Worth}(2007)}]{worth2007model}%
  \BibitemOpen
  \bibfield  {author} {\bibinfo {author} {\bibfnamefont {G.~A.}\ \bibnamefont
  {Worth}},\ }\href@noop {} {\bibfield  {journal} {\bibinfo  {journal} {J.
  Photochem. Photobiol. A}\ }\textbf {\bibinfo {volume} {190}},\ \bibinfo
  {pages} {190} (\bibinfo {year} {2007})}\BibitemShut {NoStop}%
\bibitem [{\citenamefont {Nandipati}\ and\ \citenamefont
  {Vendrell}(2021)}]{nandipati2021dynamical}%
  \BibitemOpen
  \bibfield  {author} {\bibinfo {author} {\bibfnamefont {K.~R.}\ \bibnamefont
  {Nandipati}}\ and\ \bibinfo {author} {\bibfnamefont {O.}~\bibnamefont
  {Vendrell}},\ }\href@noop {} {\bibfield  {journal} {\bibinfo  {journal}
  {Phys. Rev. Research}\ }\textbf {\bibinfo {volume} {3}},\ \bibinfo {pages}
  {L042003} (\bibinfo {year} {2021})}\BibitemShut {NoStop}%
\bibitem [{\citenamefont {Nandipati}\ and\ \citenamefont
  {Vendrell}(2020)}]{nandipati2020generation}%
  \BibitemOpen
  \bibfield  {author} {\bibinfo {author} {\bibfnamefont {K.~R.}\ \bibnamefont
  {Nandipati}}\ and\ \bibinfo {author} {\bibfnamefont {O.}~\bibnamefont
  {Vendrell}},\ }\href@noop {} {\bibfield  {journal} {\bibinfo  {journal} {J.
  Chem. Phys.}\ }\textbf {\bibinfo {volume} {153}},\ \bibinfo {pages} {224308}
  (\bibinfo {year} {2020})}\BibitemShut {NoStop}%
\bibitem [{\citenamefont {Svoboda}\ \emph {et~al.}(2022)\citenamefont
  {Svoboda}, \citenamefont {Waters}, \citenamefont {Zindel},\ and\
  \citenamefont {W{\"o}rner}}]{svoboda2022generation}%
  \BibitemOpen
  \bibfield  {author} {\bibinfo {author} {\bibfnamefont {V.}~\bibnamefont
  {Svoboda}}, \bibinfo {author} {\bibfnamefont {M.~D.}\ \bibnamefont {Waters}},
  \bibinfo {author} {\bibfnamefont {D.}~\bibnamefont {Zindel}}, \ and\ \bibinfo
  {author} {\bibfnamefont {H.~J.}\ \bibnamefont {W{\"o}rner}},\ }\href@noop {}
  {\bibfield  {journal} {\bibinfo  {journal} {Opt. Express}\ }\textbf {\bibinfo
  {volume} {30}},\ \bibinfo {pages} {14358} (\bibinfo {year}
  {2022})}\BibitemShut {NoStop}%
\bibitem [{\citenamefont {Baykusheva}\ \emph {et~al.}(2016)\citenamefont
  {Baykusheva}, \citenamefont {Ahsan}, \citenamefont {Lin},\ and\ \citenamefont
  {W{\"o}rner}}]{baykusheva2016bicircular}%
  \BibitemOpen
  \bibfield  {author} {\bibinfo {author} {\bibfnamefont {D.}~\bibnamefont
  {Baykusheva}}, \bibinfo {author} {\bibfnamefont {M.~S.}\ \bibnamefont
  {Ahsan}}, \bibinfo {author} {\bibfnamefont {N.}~\bibnamefont {Lin}}, \ and\
  \bibinfo {author} {\bibfnamefont {H.~J.}\ \bibnamefont {W{\"o}rner}},\
  }\href@noop {} {\bibfield  {journal} {\bibinfo  {journal} {Phys. Rev. Lett.}\
  }\textbf {\bibinfo {volume} {116}},\ \bibinfo {pages} {123001} (\bibinfo
  {year} {2016})}\BibitemShut {NoStop}%
\end{thebibliography}%
%merlin.mbs apsrev4-1.bst 2010-07-25 4.21a (PWD, AO, DPC) hacked
%Control: key (0)
%Control: author (8) initials jnrlst
%Control: editor formatted (1) identically to author
%Control: production of article title (-1) disabled
%Control: page (0) single
%Control: year (1) truncated
%Control: production of eprint (0) enabled
%
\end{document}